\documentclass[12pt, reqno]{amsart}

\makeatletter
\g@addto@macro{\endabstract}{\@setabstract}
\makeatother

\usepackage{graphicx}
\usepackage{amsmath, amssymb, amsthm}

\usepackage[authoryear]{natbib}

\usepackage[usenames,dvipsnames,svgnames,table]{xcolor}

\usepackage{booktabs}

\usepackage{mathrsfs}
\usepackage{bbm}

\usepackage{enumitem}
\setlist[enumerate]{itemsep=5pt,topsep=3pt}
\setlist[itemize]{itemsep=2pt,topsep=3pt}
\setlist[enumerate,1]{label=\arabic*.}

\renewcommand{\leq}{\leqslant}
\renewcommand{\geq}{\geqslant}

\usepackage{mdwlist}

\usepackage[citecolor=blue, colorlinks=true, linkcolor=blue]{hyperref}

\usepackage{caption}
\usepackage{subcaption}

\usepackage[mathscr]{euscript}

\usepackage{algorithmic}
\usepackage{algorithm}

\usepackage[left=1.2in, right=1.2in, top=1in, bottom=0.9in, includehead, includefoot]{geometry}

\DeclareMathOperator*{\argmax}{arg\,max}
\DeclareMathOperator{\interior}{int}

\DeclareMathOperator{\graph}{gr}

\newcommand{\st}{\ensuremath{\ \mathrm{s.t.}\ }}
\newcommand{\setntn}[2]{ \{ #1 : #2 \} }

\newcommand{\iidsim}{\stackrel {\textrm{ {\sc iid }}} {\sim} }

\newcommand{\dD}{\mathscr D}

\newcommand{\vV}{\mathscr V}

\newcommand{\RR}{\mathbbm R}
\newcommand{\NN}{\mathbbm N}
\newcommand{\PP}{\mathbbm P}
\newcommand{\EE}{\mathbbm E}
\newcommand{\1}{\mathbbm 1}

\newcommand{\BB}{\mathsf B}
\newcommand{\XX}{\mathsf X}
\newcommand{\YY}{\mathsf Y}
\newcommand{\ZZ}{\mathsf Z}

\newcommand{\sS}{\mathsf S}

\theoremstyle{plain}
\newtheorem{theorem}{Theorem}[section]
\newtheorem{corollary}[theorem]{Corollary}
\newtheorem{lemma}[theorem]{Lemma}
\newtheorem{proposition}[theorem]{Proposition}

\theoremstyle{definition}

\newtheorem{example}{Example}[section]

\newtheorem{assumption}{Assumption}[section]

\begin{document}

\title{}

\date{\today}

\begin{center}
  \Large Dynamic Programming with State-Dependent Discounting\footnote{We
    thank Damien Eldridge, Simon Grant, Timo Henckel, Fedor Iskhakov,
    Ruitian Lang, Andrzej Nowak, Ronald Stauber, the editor and two referees
    for many helpful comments
    and suggestions. The first author gratefully acknowledges financial
    support from ARC grant FT160100423. The second author is supported by an
    Australian Government Research Training Program (RTP) Scholarship. \\
\emph{Email:} \texttt{john.stachurski@anu.edu.au},
\texttt{junnan.zhang@anu.edu.au} }
 
  \large
    \bigskip
  John Stachurski\textsuperscript{a} and Junnan Zhang\textsuperscript{b}

  \normalsize
  \textsuperscript{a, b} Research School of Economics, Australian National University
  \par \bigskip

  \today
\end{center}

\begin{abstract} 
    This paper extends the core results of discrete time infinite horizon
    dynamic programming to the case of state-dependent discounting.  We
    obtain a condition on the discount factor process under which all of the
    standard optimality results can be recovered. We also show that the
    condition cannot be significantly weakened.  Our framework is general
    enough to handle complications such as recursive preferences and unbounded
    rewards.  Economic and financial applications are discussed.
    \vspace{1em}

    \noindent
    {\bf Keywords:} Dynamic programming; optimality; state-dependent discounting
    \\
    \noindent
    {\bf JEL\ Classification:} C61, C62
\end{abstract}

\section{Introduction}

Researchers in economics and finance routinely adopt settings where the
subjective discount rate used by agents in their models varies with the state.
For example, \cite{albuquerque2016valuation} study
an asset pricing model in which the discount rate is perturbed by an AR(1)
process. They show that the resulting demand shocks help explain the equity
premium puzzle.  \cite{mehra2002mood} find that small fluctuations in agents'
discount factors can have large effects on equity price volatility.  
\cite{schorfheide2018identifying} and \cite{gomez2020important}
likewise embed state-dependent discount factors into Epstein--Zin preferences
to generate realistic asset prices and returns.

State-dependent and time-varying discount rates are also common in studies of
savings, income and wealth.  An early example is
\cite{krusell1998income}.  In related work,
\cite{krusell2009revisiting} model the discount process as
a three state Markov chain and show how discount factor dispersion helps
their heterogeneous agent model match the wealth distribution.
\cite{fagereng2019saving} use
time-varying discount rates and portfolio adjustment frictions to explain the
positive correlation between savings rates and wealth observed in Norwegian
panel data.   
\cite*{hubmer2020sources} model
discount dynamics using a discretized AR(1) process.

State-dependent discounting is also found in analysis of fiscal and monetary
policy.  For example, \cite{eggertsson2003zero} study monetary policy in the
presence of zero lower bound restrictions with dynamic time preference shocks.
\cite{woodford2011simple} considers the government expenditure multiplier in a
similar environment.  \cite{eggertsson2011fiscal} and
\cite*{christiano2011government} study the effect of fiscal policies at the zero lower bound
on interest rates, while \cite{nakata2020equilibrium} analyze the term
structure of interest rates at the zero lower bound when agents have recursive
preferences.  In all of these models, state-dependent variation in discount
rates plays a significant role.\footnote{See also
\cite{correia2013unconventional}, \cite{hills2018fiscal},
\cite{hills2019effective} and \cite{williamson2019low}.}


In addition, state-dependent discounting is often used in studies of 
macroeconomic volatility. For example,
\cite{primiceri2006intertemporal} argue that shocks to agents' rates of
intertemporal substitution are a key source of macroeconomic fluctuations.
\cite{justiniano2008time} study the shifts in the volatility of macroeconomic
variables in the US and find that a large portion of consumption volatility
can be attributable to the variance in discount factors.  Additional research
in a similar vein can be found in \cite{justiniano2010investment},
\cite{justiniano2011investment}, \cite{christiano2014risk},
\cite{saijo2017uncertainty}, and \cite{bhandari2013taxes}.

The standard theory of dynamic programming over infinite horizons (see, e.g., 
\cite{blackwell1965discounted}, \cite{stokey1989recursive},
or \cite{bertsekas2017dynamic}) does not accommodate
state-dependent discounting.  Instead, it assumes either zero discounting (and
considers long-run average optimality) or a constant and positive discount
rate, which corresponds to a discount factor strictly less than one.  This
implies that, in the canonical setting, the Bellman operator satisfies the
conditions of Banach's contraction mapping theorem, which 
in turn provides the foundations for the standard optimality theory. 

We reconsider the standard theory when the constant discount factor $\beta$ is
replaced by a discount process $\{\beta_t\}$, so that time $t$ payoff
$\pi_t$ is discounted to present value as $\EE_z \prod_{i=0}^{t-1} \beta_i \,
\pi_t$ rather than $\beta^t \, \EE_z \, \pi_t$.  Here $z$ is the initial
condition of an exogenous Markov state process that drives evolution of the
discount factor.  We replace the traditional condition $\beta < 1$ with a
weaker ``eventual discounting'' condition: existence of a $t \in \NN$
such that $\sup_{z \in \ZZ} \EE_z \prod_{i=0}^{t-1}\beta_i < 1$.
For a finite irreducible state process, this is equivalent to existence of a
$t \in \NN$ such that $\EE \prod_{i=0}^{t-1}\beta_i < 1$, where $\EE$ is the
unconditional expectation.


We show that, when eventual discounting holds, (i) the value function satisfies the
Bellman equation, (ii) an optimal policy exists, (iii) Bellman's principle of
optimality holds, and (iv) value function iteration and
Howard policy iteration \citep{howard1960dynamic} are both convergent.   When $\beta_t$ is constant at $\beta < 1$, eventual
discounting holds at $t=1$, so these results capture the standard theory as a special
case.  

\begin{figure}
  \centering
  \includegraphics[width=0.7\textwidth]{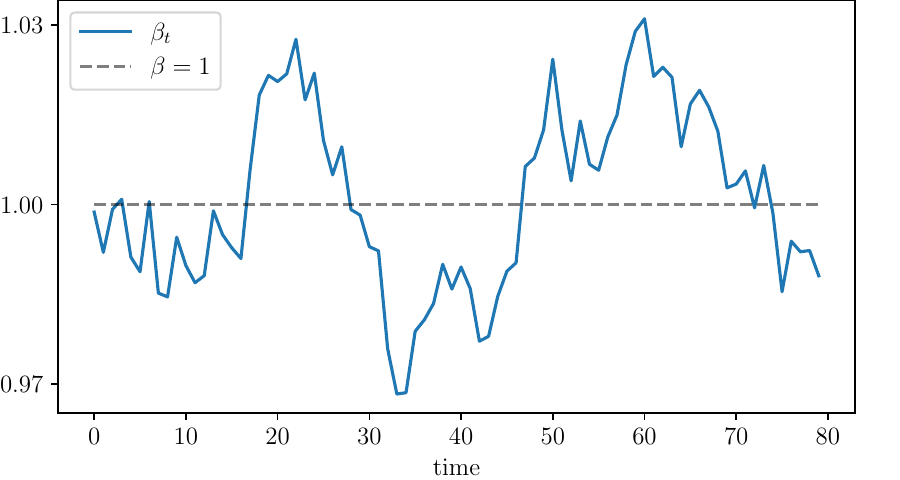}
  \caption{\label{f:hills_ar1_beta}Simulated time path for $\{\beta_t\}$ in \cite{hills2019effective}}
\end{figure}

Our conditions
do not rule out $\beta_t \geq 1$ with positive
probability. One example of why this matters is provided by the New Keynesian
literature, where the discount factor is often allowed to temporarily attain
or exceed unity, so that the zero lower bound on the nominal interest rates
binds.  For example, \citet{christiano2011government} admit a shock where
$\beta = 1.02$ in their study of the government spending multiplier.
Similarly, \citet{hills2019effective} analyze
tail risk associated with the effective lower bound on the policy rate in a
model where the discount process is a constant multiple of a discretized AR(1) process 
that regularly generates value of $\beta_t$ exceeding unity.
Figure~\ref{f:hills_ar1_beta} illustrates by showing a simulated time path of
$\{\beta_t\}$ using their parameters.\footnote{The specification is based
    around an AR(1) process and detailed in
    Example~\ref{eg:hills} below.   Other
    studies using an AR(1) specification for the discount process or
    its logarithm include \cite{nakata2016optimal},
    \cite{hubmer2020sources}, \cite{albuquerque2016valuation} and
    \cite{schorfheide2018identifying}.}

We discuss the eventual discounting condition at length in the paper, giving
several equivalent conditions.  One of these involves a bound on the spectral
radius of a discounting operator.  This connects our work to a strand of
literature in finance that study the long-term factorization of stochastic
discount factors using eigenfunctions of valuation operators (see, e.g.,
\cite{hansen2009long}, \cite{hansen2012recursive}, and \cite{qin2017long}).
Drawing on these ideas, \cite{borovivcka2020necessary} and
\cite{christensen2020existence} connect
the spectral radius of valuation operators with existence and uniqueness of
recursively defined utilities. However, neither of these papers provides
results on optimality or dynamic programming. 

To handle unbounded rewards, we extend two approaches that have been developed
previously for the case of constant discounting.  The first one treats
homogeneous programs in the spirit of \citet{alvarez1998dynamic} and
\citet[Section 9.3]{stokey1989recursive}. The second uses a local contraction
method pioneered in \cite{rincon2003existence} and further developed by
\cite{martins2010existence} and \cite{matkowski2011discounted}. 
In each case, we show how the eventual discounting condition can
be adapted to handle these extensions.

In addition, we study dynamic programming with Epstein-Zin utilities,
where rewards are unbounded above and the Bellman operator is not a
contraction in the short or long run under standard metrics. To solve the
problem we extend earlier work by \cite{marinacci2010unique},
\cite{bloise2018convex}, and \cite{becker2018recursive}, which exploits the
monotonicity and concavity of the aggregator,
to allow for state-dependent discounting.
We show that, in the case of Epstein--Zin utility, the eventual discounting condition must be adapted
to compensate for the role played by intertemporal elasticity of substitution.

Other papers have analyzed dynamic programming problems where discount
rates can vary. For example, \citet{karni2000saving} study the saving
behavior of agents with random discount factors in a steady-state
competitive equilibrium. \citet{cao2020recursive} proves the existence of
sequential and recursive competitive equilibria in incomplete markets with
aggregate shocks in which agents also have state-dependent discount factors.
In the mathematical literature, various issues in dynamic programming with
state-dependent discounting have been studied; see, for example,
\cite{jasso2020discrete} and the references
therein.\footnote{\label{fn:endogenous}\cite{jasso2020discrete} also allow the discount
  process to be endogenous, a case not covered in our framework. In economic
  applications, this often comes in the form of Uzawa type preferences
  \citep{uzawa1968time} that are common in open economy models where
  discount factors are dependent on consumption. See \cite{uribe2017open}
  for a review. However, these models can be treated using traditional
  dynamic programming techniques, since the discount factors are assumed to
  be strictly less than one in the literature.} However, these papers assume
that the discount process in the dynamic program is bounded above by
one or by some constant less than one.\footnote{\cite{schal1975conditions}
  admits state-dependent discounting in discrete time under weaker
  conditions, but he directly assumes that expected discounted rewards are
  finite under any Markov policy. This restricts all primitives in the
  dynamic program simultaneously and makes the condition impractical for
  applications.} This is too strict for many applications, as discussed
above.


Our work is related to \cite{toda2019wealth}, who investigates an income
fluctuation problem in which the agent has CRRA utility. He obtains a
necessary and sufficient condition for the existence of a solution to the
optimal saving problem with state-dependent discount factors.
\cite{ma2020income} relax the CRRA restriction by constructing optimality results
via a consumption policy operator.  
Their results are specialized to optimal savings with additively
separable rewards and do not apply to problems that involve discrete
choices, endogenous labor supply, durable goods, or other common features.  In contrast, the theory below is
developed in a general dynamic programming setting, where the state 
spaces are arbitrary metric spaces.  

In addition, the consumption policy operator, around which the theory in
\cite{toda2019wealth} and \cite{ma2020income} is constructed, is defined from
the Euler equation, which characterizes the solution in their setting.
However, many recent applications of state dependent discounting use recursive
preferences (see, e.g., \cite{albuquerque2016valuation},
\cite{basu2017uncertainty},  \cite{schorfheide2018identifying},
\cite{nakata2020equilibrium}, or
\cite{de2020valuation}), implying that the Euler equation contains the value
function and the consumption policy operator methods break down.
Our theory extends to recursive preferences
and illuminates the role of elasticity of intertemporal substitution on
eventual discounting.

The rest of this paper is structured as follows. Section~\ref{s:model} sets
out the model and provides our main results. Section~\ref{s:ex} gives
applications. Section~\ref{s:lrc} reviews our key assumption.
Sections~\ref{s:ubr} and \ref{s:ext} treat extensions. Section~\ref{s:c}
concludes.

\section{A Dynamic Program}
\label{s:model}

In what follows, for any metric space $\YY$, the symbols $m\YY$, $bm \YY$ and
$bc\YY$ denote the (Borel) measurable, bounded measurable and bounded
continuous functions from $\YY$ to $\RR$ respectively.  Unless otherwise
stated, the last two spaces are endowed with the supremum norm and this norm
is represented by $\| \cdot \|$.  In expressions with products below,
we adopt the convention that $\prod_{t=0}^{n-1} \beta_t = 1$ whenever $n=0$.

\subsection{Framework}

\label{ss:fr}

The state of the world consists of a pair $(x, z)$, where $x$ and $z$ represent 
endogenous and exogenous variables.  These variables take
values in separable metric spaces $\XX$ and $\ZZ$ respectively.
The agent responds to $(x,z)$ by choosing future state $x'$ 
from $\Gamma(x, z) \subset \XX$, where $\Gamma$ is the \emph{feasible
correspondence}.  Let $\graph \Gamma$ be the graph of $\Gamma$, defined by
\begin{equation}
    \graph \Gamma = \setntn{ (x, z, x') \in \sS \times \XX} { x' \in \Gamma(x, z)}   
    \quad
    \text{ where $\sS := \XX \times \ZZ$}. 
\end{equation}
Similar to~\cite{bertsekas2013abstract}, we combine the remaining elements of
the dynamic programming problem into a single \emph{continuation aggregator}
$H$, with the understanding that $H(x, z, x', v)$ is the maximal value that can
be obtained from the present time under the continuation value function $v$,
given current state $(x, z)$ and next period state $x'$.  The aggregator $H$ maps
each $(x, z, x', v)$ in $\graph \Gamma \times bm\sS$ into $\RR$ and is assumed
to satisfy,  for all $v, w\in bm\sS$  and all $(x, z, x') \in \graph \Gamma$,   
\begin{equation}\label{eq:mon}
    H(x, z, x', v) \leq H(x, z, x', w) \text{ whenever } v \leq w.
\end{equation}
This basic monotonicity
condition is satisfied in all applications of interest.
Bellman's equation takes the form
\begin{equation}
  \label{eq:bellman0}
  v(x, z) = \sup_{x'\in \Gamma(x, z)} H(x, z, x', v).
\end{equation}

For fixed $\XX$ and $\ZZ$, a \emph{dynamic
program} $\dD = (\Gamma, H)$ consists of a feasible correspondence $\Gamma$ and a
continuation aggregator $H$.


\subsection{Feasibility and Optimality}\label{ss:feasop}

Let $\dD = (\Gamma, H)$ be a dynamic program and let $\Sigma$ be the set of
\emph{feasible policies}, defined as all Borel measurable maps $\sigma$ from
$\sS$ to $\XX$ such that $\sigma(x, z) \in \Gamma(x, z)$ for each $(x, z)$ in
$\sS$.  Given such $\sigma$, let $T_\sigma$ be the \emph{policy operator} on
$bm\sS$ given by 
\begin{equation}
  \label{eq:T_sigma}
  (T_\sigma v)(x, z) = H(x, z, \sigma(x, z), v).
\end{equation}
Define the \emph{Bellman operator} $T$ on $bm\sS$ by
\begin{equation}
  \label{eq:bellman}
  (T v)(x, z) = \sup_{x'\in \Gamma(x, z)} H(x, z, x', v).
\end{equation}
Given $v_0$ in $bm\sS$ and $\sigma$ in $\Sigma$, we can interpret
  $v_{n, \sigma}(x, z) := (T_\sigma^n v_0)(x, z)$
as the lifetime payoff of an agent who starts at state $(x, z)$,
follows policy $\sigma$ for $n$ periods and uses $v_0$ to evaluate the
terminal state.  The 
\emph{$\sigma$-value function} for an infinite-horizon problem is defined here
as 
\begin{equation}\label{eq:defvs}
  v_\sigma(x, z) := \lim_{n\to\infty} v_{n, \sigma}(x, z).
\end{equation}
The definition requires that this
limit exists and is independent of $v_0$.  Below we impose
conditions such that this is always the case.

We define the \emph{value function} corresponding to our dynamic program
by
\begin{equation}
    \label{eq:dvf}
    v^*(x, z) = \sup_{\sigma\in\Sigma} v_\sigma(x, z)
\end{equation}
at each $(x,z)$ in $\sS$.  A policy $\sigma^* \in \Sigma$ is called
\emph{optimal} if it attains the
supremum in \eqref{eq:dvf}
at each $(x, z)$ in $\sS$.
We say that \emph{Bellman's principle of optimality holds} when
\begin{equation*}
    \sigma \in \Sigma \text{ is optimal }
    \;\iff\;
  \sigma(x, z) \in \argmax_{x'\in \Gamma(x, z)} H(x, z, x', v^*)
  \, \text{ for each $(x, z)$ in $\sS$}.
\end{equation*}

\subsection{Assumptions}
\label{ss:reg}
 
A dynamic program $\dD = (\Gamma, H)$ will be called \emph{regular} if
\begin{enumerate}[label=(\alph*)]
    \item $\Gamma$ is continuous, nonempty,
        and compact valued and
    \item the function $(x, z, x') \mapsto H(x, z, x', v)$ is bounded and measurable
        on $\graph \Gamma$ for all $v\in bm\sS$, and also 
        continuous when $v\in bc\sS$.
\end{enumerate}
Most standard cases from the literature are regular, 
including all dynamic programs with a finite state space.\footnote{\label{fn:cont}The
continuity and compactness conditions are automatically satisfied when $\XX$
and $\ZZ$ are finite and endowed with the discrete topology.} Further
discussion of regularity is provided in Section~\ref{s:ex}.

Let $\beta_t = \beta(Z_t) \geq 0$ for some $\beta \in bm\ZZ$ 
and Markov process $\{Z_t\}$ on $\ZZ$ with transition kernel $Q$.\footnote{That is, $Q(z, B) = \PP\{Z_{t+1} \in B \,|\, Z_t = z\}$
for all $z \in \ZZ$ and $B$ in the Borel subsets of $\ZZ$.}
Let $\EE_z$ represent expectation given $Z_0 = z$.
We call $(\beta, Q)$ \emph{eventually discounting} if $r_n^\beta < 1$ for some $n
\in \NN$, where
\begin{equation*}
    r_n^\beta := \sup_{z\in \ZZ} \, \EE_z \, \prod_{t=0}^{n-1} \beta_t.
\end{equation*}

\begin{example}
    If there exists a constant $b \geq 0$ such that
    $\beta_t \equiv b$ for all $t \geq 0$, then $r_n^\beta = b^n$.
    Eventual discounting holds if and only if $b < 1$.   
\end{example}

\begin{example}
    If $\{Z_t\}$ is {\sc iid}, then $r_n^\beta =
    \prod_{t=0}^{n-1} \EE \beta_t =
    b^n$ where $b := \EE \beta_t$.  Hence eventual discounting holds if and
    only if $\EE \beta_t < 1$.  In particular, higher moments have no
    influence on eventual discounting unless there is persistence.
\end{example}

Section~\ref{s:lrc} provides
an extended discussion of eventual discounting for more sophisticated state
processes.
  
\begin{assumption}[Eventual Contractivity]\label{a:con}
    There is a nonnegative function $\beta$ in $bc\ZZ$ and a Feller
    transition kernel $Q$ on $\ZZ$ such that $(\beta, Q)$ is eventually
    discounting and
    \begin{equation}
      \label{eq:bc}
      |H(x, z, x', v) - H(x, z, x', w)| 
         \leq \beta(z)
         \int |v(x', z') - w(x', z')|Q(z, dz')
    \end{equation}
    for all $v, w\in bm\sS$ and $(x, z, x') \in \graph
    \Gamma$.\footnote{Here we implicitly assume that the discount factor is
      known to the agent at the beginning of each period. Our results hold
      for alternative timing with slight modifications to~\eqref{eq:bc}. See
      Section~\ref{ss:alt}.}
\end{assumption}

The Feller property means that either $\ZZ$ is
discrete or the law of motion is continuous.\footnote{\label{fn:feller}More precisely, we
    assume that, for any $h \in bc\sS$, the function 
    $(x, z) \mapsto \int h(x, z') Q(z, d z')$ is
    continuous.  This holds 
automatically when $\ZZ$ is countable (under the discrete topology).
It also holds if $Q$ is generated by a continuous law of motion, in the sense
that $Z_{t+1} = F(Z_t, W_{t+1})$ for some
continuous function $F$ and {\sc iid} sequence $\{W_t\}$.
These two cases cover all the applications we consider.  Further discussion
can be found in Lemma~12.14 of \cite{stokey1989recursive}.}

\subsection{Optimality Results}\label{ss:opt}

In the statement of the next theorem,
 a map $M$ from a metric
space into itself is called \emph{eventually contracting} if there exists an
$n$ in $\NN$ such that the $n$-th iterate $M^n$ is a contraction mapping.\footnote{More
    precisely, a self-map $M$ on metric
space $(Y, \rho)$ is called eventually contracting if there
exists an $n$ in $\NN$ and a $\lambda < 1$ such that $\rho(M^n y, M^n y') \leq
\lambda \rho(y, y')$ for all $y, y'$ in $Y$.}

\begin{theorem}
    \label{t:bk}
    Let $\dD$ be a dynamic program.
    If $\dD$ is regular and Assumption~\ref{a:con} holds, then the
    following statements are true:
    \begin{enumerate}[label=\textnormal{(\alph*)}]
        \item $T_\sigma$ is eventually contracting on $bm\sS$ and $T$ is eventually contracting on $bc\sS$.\label{bk-a}
        \item For each feasible policy $\sigma$, the lifetime value
          $v_\sigma$ is a well defined element of $bm\sS$.\label{bk-b}
        \item The value function $v^*$ is finite, continuous, and the only
          fixed point of $T$ in $bc\sS$.\label{bk-c}
        \item At least one optimal policy exists.\label{bk-d}
        \item Bellman's principle of optimality holds.\label{bk-e}
    \end{enumerate}
    In addition, value function and Howard policy iteration converge:
    \begin{enumerate}[resume, label=\textnormal{(\alph*)}]
        \item $\lim_{k \to \infty} T^k v = v^*$ for all $v \in bc\sS$ and \label{bk-f}
        \item $\lim_{k \to \infty} v_{\sigma_k} = v^*$ when
            $\sigma_k \in \Sigma$ and
                $T_{\sigma_k} v_{\sigma_{k-1}} = T v_{\sigma_{k-1}}$
                 for all $k\in \NN$.\label{bk-g}
    \end{enumerate}
\end{theorem}

This theorem extends the core results of dynamic programming
theory to the case of state-dependent discounting.  In
particular, the value function satisfies the Bellman equation, an optimal
policy exists, and Bellman's principle of optimality is valid.
Value iteration and policy iteration both lead to the value function,
so that we have both existence of an optimal policy and means to compute it.
The proof of Theorem~\ref{t:bk} can be found in the appendix.

Relative to the results that can be obtained under standard
contraction conditions (see, e.g., \cite{bertsekas2013abstract}), the only significant weakening of
the main findings is that $T$ and $T_\sigma$ are eventually contracting,
rather than always contracting in one step.
Such an outcome cannot be avoided when values of the discount factor greater than one
are admitted.


The eventual discounting condition is, in many cases, not just sufficient but
also necessary for the dynamic program to be well defined and the optimality
results to hold.  
Appendix~\ref{ss:nec} provides additional discussion.

\subsection{Blackwell's Condition}

Blackwell's sufficient condition for a contraction has a
  natural analogue in the case of state-dependent discounting. As shown in
Proposition~\ref{p:blackwell}, if the Bellman operator satisfies
\begin{equation*}
    [T(v+c)](x, z) \leq (Tv)(x, z) + \beta(z)\int c(z')Q(z, dz')
    \qquad ((x, z) \in \sS)
\end{equation*}
for all $c \in bm\ZZ_+$ where $(\beta, Q)$ is eventually discounting, then $T$ is
eventually contracting on $bc\sS$.  As a consequence, $T$ has a unique fixed
point in $bc\sS$ that is globally attracting under iteration of $T$.  This
extends Blackwell's original result,\footnote{The original
  result states that if an operator $T$ is monotone and there exists a $b\in
  (0, 1)$ such that $T(v+c) \leq Tv + bc$ for all $c \geq 0$, then $T$ is a
  contraction \cite[see, e.g.,][Theorem 3.3]{stokey1989recursive}.} with the caveat that $T$ might not
itself be a contraction.  Again, this cannot be avoided when $\beta$ is
allowed to take values greater than one.\footnote{In fact, when $T$ is an
    eventual contraction on a Banach space, one can construct a complete metric on the same
    space under which $T$ is a contraction.  See, for example,
    \cite{krasnoselskii1972approximate}.  Our terminology on contractions in
    this section refers specifically to the supremum norm.}

\subsection{Monotonicity, Concavity and Differentiability}\label{ss:mon}

Next we show that standard results on monotonicity, concavity, and
differentiability of the value function (cf, e.g.,
\cite{stokey1989recursive}) are preserved under state-dependent
discounting without additional assumptions on the discount factor
process. We assume that $\XX$ is a convex subset of $\RR$ in the
discussion below and denote $ibc\sS$ the set of functions in $bc\sS$ that are
increasing and concave in $x$.

\begin{assumption}
  \label{a:ibc}
  For all $v\in ibc\sS$ and $z\in\ZZ$, (i) $x \mapsto H(x, z, x', v)$ is
  increasing for all $x'\in \Gamma(x, z)$, (ii) $(x, x') \mapsto H(x, z, x',
  v)$ is strictly concave, (iii) $\Gamma(x, z) \subset \Gamma(y, z)$ for all
  $x \leq y$, and (iv) the set $\left\{(x, x'): x'\in\Gamma(x, z)\right\}$
  is convex.
\end{assumption}

\begin{assumption}
  \label{a:diff}
  The map $x \mapsto H(x, z, x', v)$ is continuously differentiable on
  $\interior\XX$ for all $z\in\ZZ$, $x'\in\interior \Gamma(x, z)$, and $v\in
  ibc\sS$.
\end{assumption}

The following theorem shows that the value function $v^*$ is increasing,
strictly concave, and continuously differentiable in $x$ under standard
assumptions.\footnote{If $\dD$ is additively separable, sufficiency
    of the Euler equations and transversality conditions can also be
    established, analogous to Section 9.5 of \cite{stokey1989recursive}.}

\begin{theorem}
  \label{t:mon}
  If $\dD$ is regular and
  Assumptions~\ref{a:con}--\ref{a:ibc} hold, then $x \mapsto v^*(x, z)$
  is increasing and strictly concave and $x \mapsto \sigma^*(x, z)$ is
  single-valued and continuous for all $z\in\ZZ$.  If, in addition,
  Assumption~\ref{a:diff} holds, then $x \mapsto v^*(x, z_0)$ is continuously
  differentiable at $x_0$ whenever $x_0\in\interior\XX$ with $\sigma^*(x_0,
  z_0)\in \interior\Gamma(x_0, z_0)$ for some $z_0$,
  and
  \begin{equation*}
    v^*_x(x_0, z_0) = H_x(x_0, z_0, \sigma^*(x_0, z_0), v^*).
  \end{equation*}
\end{theorem}

Additional comments on these assumptions and results can be found in the
applications.

\section{Additively Separable Problems}\label{s:ex}

In this section we study state-dependent discounting in settings
where preferences are additively separable and rewards are bounded.
(Extensions to unbounded rewards and recursive preferences are deferred to
Sections~\ref{s:ubr} and \ref{s:ext}.) 

\subsection{An Additively Separable Problem} \label{ss:genadsep}

Consider the dynamic program in Section~9.2 of~\cite{stokey1989recursive} with
the addition of state-dependent discounting.  The objective is to maximize
\begin{equation}
  \label{eq:ogo}
      \EE 
          \sum_{t=0}^\infty 
          \, \prod_{i=0}^{t-1} \beta_i \, F(X_t, Z_t, X_{t+1})
      \quad
  \text{s.t. $X_{t+1} \in \Gamma(X_t, Z_t)$ for all $t \geq 0$}.
\end{equation}
As in \cite{stokey1989recursive}, $F$ is assumed to be bounded and
continuous on $\graph \Gamma$, while $\Gamma$ is a continuous, nonempty,
and compact-valued.  
We set $\beta_t = \beta(Z_t)$ where $\beta$ is continuous, bounded and nonnegative,
while $\{Z_t\}$ is Markov with Feller kernel $Q$.

We connect this dynamic program to our framework by setting $\dD = (\Gamma,
H)$ with 
\begin{equation}
  \label{eq:ogo3}
  H(x, z, x', v) := F(x, z, x') + \beta(z) \int v(x', z')Q(z, dz')
\end{equation}
for all $v \in bm\sS$.  
The monotonicity condition~\eqref{eq:mon} is clearly satisfied. 
The function $(x, z, x') \mapsto H(x, z, x', v)$ is bounded and
Borel measurable on $\graph \Gamma$ because $v$ and $F$ have these properties,
and continuous when $v$ is continuous by the Feller property (see footnote~\ref{fn:feller}).  Hence $\dD$ is regular.  

If $(\beta, Q)$ is eventually discounting
then Assumption~\ref{a:con} holds, since \eqref{eq:ogo3} yields
\begin{equation*}
  |H(x, z, x', v) - H(x, z, x', w)| 
         \leq \beta(z)
         \left|
         \int [v(x', z') - w(x', z')] Q(z, dz')
         \right|,
\end{equation*}
and an application of the triangle inequality gives \eqref{eq:bc}.

To connect this application with the definition of optimality given in
Section~\ref{ss:feasop}, fix $\sigma \in \Sigma$ and $v \in bm\sS$. 
The policy operator $T_\sigma$ from \eqref{eq:T_sigma} can be expressed as
\begin{equation}
  \label{eq:poli}
    (T_\sigma v)(x_0, z_0) =  F(x_0, z_0, \sigma(x_0, z_0)) 
    + \beta(z_0) \, \EE_0 \, v(X_1, Z_1)
\end{equation}
where $\{X_t\}$ is generated by $X_{t+1} = \sigma(X_t, Z_t)$, the initial
condition is $(X_0, Z_0) = (x_0, z_0)$, and $\EE_t$
conditions on $\{Z_i\}_{i \leq t}$.
If we take $T_\sigma$, iterate forward $n$ times and apply the law of
iterated expectations, we obtain
\begin{equation}\label{eq:tnsls}
    (T^n_\sigma v)(x_0, z_0)
    = \EE_0 \, \sum_{t=0}^{n-1} 
    \prod_{i=0}^{t-1} \beta_i F(X_t, Z_t, X_{t+1})
    + \EE_0 \, \prod_{i=0}^{n} \beta_i v(X_n, Z_n).
\end{equation}
Recall from~\eqref{eq:defvs} that, to obtain the value $v_\sigma$ of
the policy $\sigma$, we take the limit of \eqref{eq:tnsls} in $n$.
Eventual discounting implies that the second
term vanishes as $n \to \infty$.\footnote{This term is
    dominated by $r_{n+1}^\beta \, \|v \| $.  Hence it suffices to prove that
    $r^\beta_n \to 0$
    as $n \to \infty$.  Eventual discounting implies that $r^\beta_n < 1$ for some $n$, and, as shown in 
    Proposition~\ref{p:0} below, this in turn gives $\lim_{n \to \infty}
    (r^\beta_n)^{1/n}
< 1$.  But then $r^\beta_n \to 0$, as was to be shown.}
    In the
limit we obtain as $v_\sigma$ the value in \eqref{eq:ogo} under the policy
$\sigma$.  Maximizing over $\sigma$ in $\Sigma$ yields the optimal policy.

The Bellman operator corresponding to $\dD$ is the map $T$ on $bc\sS$ defined
by %
\begin{equation}
  \label{eq:ogo2}
  (Tv)(x, z) = \max_{x' \in \Gamma(x, z)}
  \left\{ F(x, z, x') + \beta(z) \int v(x', z') Q(z, dz') \right\}.
\end{equation}
Since the conditions of Theorem~\ref{t:bk} are satisfied, the unique fixed
point of $T$ in $bc\sS$ is $v^* := \sup_{\sigma \in \Sigma} v_\sigma$, the value
function of $\dD$.  Bellman's principle of optimality applies and an optimal
policy can be computed by either value function iteration or Howard's policy
iteration algorithm.  Monotonicity, concavity and differentiability of $v^*$ can be
obtained by imposing the same conditions that \cite{stokey1989recursive}
impose on $F$ and $\Gamma$ and then applying Theorem~\ref{t:mon}.

\subsection{Application to a Savings Problem}\label{ss:saving}

The dynamic program associated with the household problem in
\cite{hubmer2020sources} can be placed with the framework provided in the
previous section.  The continuation aggregator takes the form
\begin{equation}
  \label{eq:oghug}
  H(x, z, x', v) = u(R (x, z) x + y(x, z) - x') + \beta(z) \int v(x', z')Q(z, dz')
\end{equation}
where $x \in \XX := \RR_+$ is current assets, $z$ is a vector of exogenous shocks taking
values in $\RR^k$, $R(x, z)$ is the gross rate of
return on asset holdings (which depends on both exogenous shocks and current
asset holdings) and $y(x, z)$ is labor income net of
income tax and capital gains tax, as well as a lump sum transfer. 
The utility function is
\begin{equation}
    u(c) := \frac{c^{1-\gamma}}{1-\gamma} \text{ where } \; \gamma > 1.
\end{equation}
Next period assets $x'$ are constrained to lie in 
\begin{equation}\label{eq:hugc}
    \Gamma(x, z)
    := \setntn{x' \in \RR}{\bar x \leq x' \leq R (x, z) x + y(x, z)}.
\end{equation}

This problem is not regular because $H$ is not bounded, since $u$ is unbounded below.
However, in solving this dynamic program, \cite{hubmer2020sources} reduce both the
asset space $\XX$ and the exogenous shock space $\ZZ$ to a finite grid.
The aggregator is then bounded and the continuity parts of the regularity
condition are automatically satisfied (under the discrete topology).
Hence, to show that all of the conclusions of Theorem~\ref{t:bk} apply, we
need only verify that eventual discounting holds. This issue is discussed
for the parameterization in \cite{hubmer2020sources} in Section~\ref{s:lrc}
below.

\section{The Discount Condition}\label{s:lrc}

In this section we discuss tests for the eventual discounting condition 
and develop intuition regarding its value. 

\subsection{Connection to Spectral Radii}

Given $\beta$ and $Q$ as in Assumption~\ref{a:con}, let $L_\beta \colon bm\ZZ
\to bm\ZZ$ be the \emph{discount operator} defined by
\begin{equation}
  \label{eq:B}
  (L_\beta h) (z) = \beta(z) \int h(z') Q(z, dz') 
  \qquad (h \in bm\ZZ, \; z \in \ZZ).
\end{equation}
The next proposition shows that we can 
test Assumption~\ref{a:con} by computing the
spectral radius $r(L_\beta)$ of the operator $L_\beta$.\footnote{As usual, the spectral radius of a
        bounded linear operator $L$ from a Banach space $\BB$ to itself is given by $r(L) 
        := \lim_{n\to\infty} \|L^n\|^{1/n}$, where $\|\cdot\|$ is the
        operator norm.  This limit always exists and is equal to 
        $\inf_{n\in \NN} \|L^n\|^{1/n}$.
        If $\BB$ is finite dimensional, it equals the
        maximal modulus of the eigenvalues of $L$.  See, for example,~\cite{buhlerfunctional}, Theorem~1.5.5.}
In stating it, we set $\beta_t := \beta(Z_t)$ where
$\{Z_t\}$ is a $\ZZ$-valued Markov process generated by $Q$.

\begin{proposition}\label{p:0}
    The spectral radius of $L_\beta$ satisfies
    $r(L_\beta) = \lim_{n\to\infty}  (r_n^\beta)^{1/n}$.
    Moreover, $(\beta, Q)$ is eventually discounting if and only if $r(L_\beta) < 1$.
\end{proposition}

The expression for $r(L_\beta)$ in Proposition~\ref{p:0} is obtained through a local spectral
radius condition for positive linear operators.  It provides both a simple
representation of the spectral radius of $L_\beta$ and a link to eventual
discounting.  For example, it is immediate from $r(L_\beta) =
\lim_{n\to\infty}  (r_n^\beta)^{1/n}$ that $r^\beta_n \to
0$ when $r(L_\beta) < 1$. This, in turn, implies that 
$(\beta, Q)$ is eventually discounting.  The converse implication is more
subtle and involves the Markov property.  Details are in the appendix.

\subsection{Finite Exogenous State}\label{ss:finite}

Testing eventual discounting is simple when $\ZZ$ is finite. In
this case, $Q$ can be represented as a Markov matrix of values $Q_{ij}$,
giving the one-step probability of transitioning from $z_i$ to $z_j$, and
$L_\beta$ can be represented as the matrix 
\begin{equation}\label{eq:lbm}
    L_\beta := \left( \beta_i Q_{ij} \right)_{1 \leq i, j \leq N}   .
\end{equation}
Here $\beta_i := \beta(z_i)$ and $N$ is the
number of elements in $\ZZ$.  The spectral radius $r(L_\beta)$ is equal to the
dominant eigenvalue of $L_\beta$, which is real and
    nonnegative by the Perron--Frobenius Theorem.  In view of Proposition~\ref{p:0}, eventual
discounting holds if and only if this eigenvalue is strictly less than
unity.

\begin{example}
    \cite{christiano2011government} consider the case $\beta_t \in 
    \{ \beta^\ell,  \beta^h\}$ with $\beta^\ell < 1 < \beta^h$. 
    The process $\{\beta_t\}$ stays at $\beta^h$ 
    with probability $p$ and shifts
    permanently to $\beta^\ell$ with probability $1-p$. Thus, by \eqref{eq:lbm},
    \begin{equation*}
      L_\beta = 
        \begin{pmatrix}
          \beta^\ell & 0\\
          (1-p)\beta^h & p\beta^h
        \end{pmatrix}.
    \end{equation*}
    The eigenvalues  are $\beta^\ell$ and
    $p\beta^h$, so $r(L_\beta)$ is the maximum of these values.  
    Since $\beta^\ell < 1$, eventual discounting holds if and only if 
    $p\beta^h < 1$.  The condition is violated if the state $\beta^h$ is too 
    large or too persistent. \cite{christiano2011government} set $\beta^h =
    1.02$ and consider $p \leq 0.82$, so eventual discounting is satisfied.
\end{example}

\subsection{Stationary Spectral Radius}

The expression obtained for $r(L_\beta)$ in Proposition~\ref{p:0} is a geometric
mean, and hence is determined by the asymptotic behavior of the discount process. When $\{Z_t\}$ is
irreducible, it seems likely that these asymptotics will be independent of
the initial condition $z$. This suggests that the conditional expectation
and supremum in the definition of $r^\beta_n$ can be replaced by the unconditional
expectation $\EE$ for the stationary process. The next proposition confirms
this intuition.

\begin{proposition}\label{p:fic}
    If $\ZZ$ is finite and the exogenous state process $\{Z_t\}$ is irreducible, then
    $r(L_\beta)$ satisfies the stationary representation
    \begin{equation}\label{eq:ssr}
        r(L_\beta) = s^\beta
        \quad \text{where} \quad
        s^\beta := \lim_{n\to\infty}  (s^\beta_n)^{1/n} 
        \quad \text{with} \quad
        s^\beta_n := \EE \prod_{t=0}^{n-1}\beta_t.
    \end{equation}
\end{proposition}

Our analysis below shows that this stationary representation is also highly
accurate even when $\ZZ$ is infinite, provided that $\{Z_t\}$ is irreducible
and sufficiently mean reverting for dependence on initial conditions to die
out.  This is helpful because the stationary representation
of $r(L_\beta)$ sometimes admits analytical solutions that facilitate
benchmark calculations and enhance intuition.\footnote{While finiteness of the
    state space can be weakened, as discussed above, irreducibility is
    essential.  To see this, consider the application in 
  \cite{christiano2011government},
  where the unique stationary distribution puts all mass on the low
  state and irreducibility fails. With all mass on the low state we have
  $s^\beta_n
  = (\beta^\ell)^n$ for all $n$, and hence $s^\beta=\beta^\ell$, which differs
  from $r(L_\beta) = \max\{\beta^\ell, p \beta^h\}$.}

\subsection{Autoregressive Specifications}\label{ss:ar}

Some studies adopt discount processes that are autoregressive in
levels or logs \citep[e.g.,][]{hubmer2020sources, hills2019effective,
  nakata2016optimal} and then discretize them prior to computation. Such 
  specifications always fit the dynamic programming framework adopted above after discretization.\footnote{Recall that
  $\beta$ is assumed to be bounded and continuous in
  Assumption~\ref{a:con}.  Both conditions hold after discretization.
  (Continuity holds automatically under the discrete topology.)}  The only
  remaining issue is whether or not eventual discounting holds.  For common
  reference, all examples use the state process
\begin{equation}
    \label{eq:ar1}
    Z_{t+1} = \rho Z_t + (1-\rho)\mu + \sigma_\epsilon \epsilon_{t+1},
    \quad \{\epsilon_t\} \iidsim N(0, 1).
\end{equation}

\subsubsection{AR(1) in Levels}

We first give examples where $\beta_t$ is a multiple of $Z_t$.  After following the discretization procedure used by the
  authors, we calculate the spectral radius of the matrix~\eqref{eq:lbm}.


\begin{example}\label{ex:hubmer}
    \cite{hubmer2020sources} take the AR(1) specification $\beta_t = Z_t$
    where $\{Z_t\}$ follows~\eqref{eq:ar1}
    with $\rho = 0.992$, $\mu = 0.944$ and $\sigma_\epsilon = 0.0006$
    and discretize the process onto a grid of 15 states via Tauchen's method.
    This gives $r(L_\beta) = 0.9469$, so eventual discounting holds.
    This is as expected, since the mean $\mu$ is substantially less than one
    and low volatility suggests that the impact of stochastic variation is minor.
\end{example}

\begin{example}\label{eg:hills}
    In \cite{hills2019effective}, the discount process is $\beta_t =
    b Z_t$ where $\{Z_t\}$ obeys \eqref{eq:ar1}.  They consider several
    parameterizations, the most empirically motivated of which is 
    $\mu=1$, $b = 0.99875$, $\rho = 0.85$ and
    $\sigma_\epsilon = 0.0062$.  Under this parameterization
    $\beta_t$ regularly exceeds one, as observed in the simulated
    process shown in Figure~\ref{f:hills_ar1_beta}.
    Nonetheless, after following their discretization procedure and
    computing the spectral radius of $L_\beta$, we find $r(L_\beta)=0.9996$,
    so eventual discounting holds.
\end{example}

\begin{example}
    In a similar setting to Example~\ref{eg:hills}, \cite{nakata2016optimal} assumes $\beta_t = bZ_t$ where
    $\{Z_t\}$ follows \eqref{eq:ar1}, $\mu = 1$, $b = 0.995$, $\rho = 0.85$,
    and $\sigma_\epsilon = 0.00395$. The process is discretized onto a grid of 501
    points, yielding $r(L_\beta) = 0.9953$.
\end{example}

\begin{figure}
    \centering
    \includegraphics[trim={0 0 0 3ex},width=0.7\textwidth]{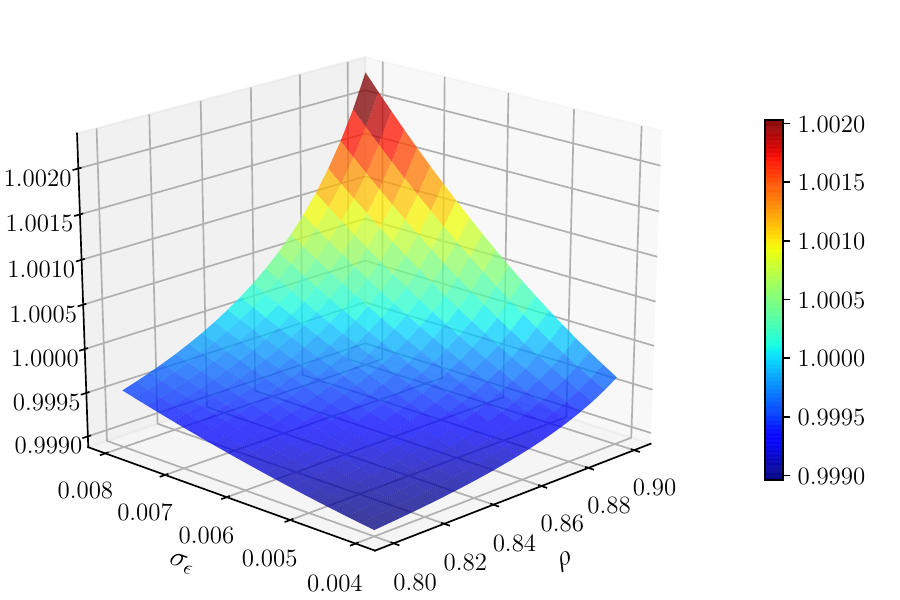}
    \caption{$r(L_\beta )$ as a function of $\rho$ and $\sigma_\epsilon$; $\mu=0.944$}
    \label{fig:r}
\end{figure}

To illustrate how the stochastic properties of $\beta_t$ affect the
size of $r(L_\beta)$, we take the parameterization in Example~\ref{eg:hills} as a benchmark
and vary the persistence term $\rho$ and the volatility $\sigma_\epsilon$.
Other parameters are held constant.
Figure~\ref{fig:r} plots the resulting values of $r(L_\beta)$.
The figure shows that
higher volatility and higher persistence both increase $r(L_\beta)$, leading to
a failure of eventual discounting when $r(L_\beta) \geq 1$.
Note also that there is a positive interaction between persistence and
volatility, with the effect of each parameter enhanced by the other.

Some further insight can be gained by considering the expected two period
discount factor when $\beta_t = Z_t$ and $\{Z_t\}$ is as given
in~\eqref{eq:ar1}.  Under the stationary distribution, which governs
asymptotic outcomes, this evaluates to 
\begin{equation}\label{eq:bpro}
    \EE \beta_t \beta_{t+1} 
    = \mu^2 + \rho \frac{\sigma_\epsilon^2}{1-\rho^2} .
\end{equation}
The value in~\eqref{eq:bpro} depends on the sign of $\rho$.
Positive correlation combined with positive volatility in the state process
leads to a value greater than the stationary mean.
This is because, under positive correlation, positive deviations from the mean
tend to occur consecutively and reinforce each other.

\subsubsection{AR(1) in Logs}\label{sss:ar1logs}

Next we set $\beta_t := \exp(Z_t)$ where $\{Z_t\}$ obeys the AR(1)
specification \eqref{eq:ar1}.
This specification is
arguably more natural than the direct AR(1) approach
discussed above due to positivity. 
While the state space is not finite,
irreducibility of $\{Z_t\}$ leads us to conjecture that
an approximate version of Proposition~\ref{p:fic} holds, 
so that the stationary geometric mean $s^\beta = \lim_{n\to \infty}
(s^\beta_n)^{1/n}$
for the original process will be close to $r(L_\beta) = \lim_{n\to \infty}
(r^\beta_n)^{1/n}$
when the latter is calculated using an appropriately discretized version of the process.  As shown in Appendix~\ref{s:geomean}, for
the original process we have
\begin{equation}
  \label{eq:geomean}
  s^\beta 
  = \lim_{n \to\infty} (s^\beta_n)^{1/n}
  = \lim_{n \to\infty} 
      \left(\EE \prod_{t=0}^{n-1} \beta_t \right)^{1/n} 
      = \exp \left\{\mu + \frac{\sigma_\epsilon^2}{2(1-\rho)^2} \right\}.
\end{equation}
Numerical experiments show that the expression on the right hand side
of~\eqref{eq:geomean} provides a good approximation of
$r(L_\beta)$ even when the discretization is relatively coarse, and an almost
perfect approximation when the discretization is fine.
Table~\ref{tab:r} illustrates by comparing $s^\beta$ given by
\eqref{eq:geomean} and $r(L_\beta)$ under two different levels of
discretization, for a range of parameter values.\footnote{$N$ is the number of grid points. We use the Rouwenhorst's method
  for discretization, which has strong asymptotic properties in
  terms of approximating the distributions of Gaussian AR(1) processes
  \citep{kopecky2010finite}. We fix $\mu$ because it has no effect
  on the errors.}

\begin{table}[tb!]
    \caption{Comparison of $s^\beta$ and $r(L_\beta)$ after discretization}
    \label{tab:r}
    \footnotesize
    \begin{tabular*}{\textwidth}{@{} l @{\extracolsep\fill} c
      @{\extracolsep\fill} c @{\extracolsep\fill} c @{\extracolsep\fill} c
      @{\extracolsep\fill} c @{}}
      \toprule
      Parameters &  & \multicolumn{2}{@{}l}{N=10} & \multicolumn{2}{@{}l}{N=200}\\
      \cmidrule{3-4} \cmidrule{5-6}
      $\mu = -0.05$ & $s^\beta$ & $r(L_\beta)$ & Error & $r(L_\beta)$ & Error \\
      \midrule
      $\rho=0.90,\;\sigma_\epsilon=0.01$ & 0.956 & 0.956 & 2.5e-05 & 0.956 & 1.1e-06 \\
      $\rho=0.90,\;\sigma_\epsilon=0.02$ & 0.970 & 0.970 & 3.9e-04 & 0.970 & 1.8e-05 \\
      [1ex]
      $\rho=0.92,\;\sigma_\epsilon=0.01$ & 0.959 & 0.959 & 7.6e-05 & 0.959 & 3.5e-06 \\
      $\rho=0.92,\;\sigma_\epsilon=0.02$ & 0.981 & 0.980 & 1.2e-03 & 0.981 & 5.8e-05 \\
      [1ex]
      $\rho=0.94,\;\sigma_\epsilon=0.01$ & 0.965 & 0.964 & 3.2e-04 & 0.965 & 1.5e-05 \\
      $\rho=0.94,\;\sigma_\epsilon=0.02$ & 1.006 & 1.001 & 4.7e-03 & 1.005 & 2.5e-04 \\
      \bottomrule
    \end{tabular*}
\end{table}

Given this tight relationship between $s^\beta$ and $r(L_\beta)$, we can use
\eqref{eq:geomean} to examine how 
the parameters of the state process affect eventual discounting.
Consistent with our previous findings,
the expression in~\eqref{eq:geomean} indicates that 
$r(L_\beta)$ is increasing
in all of the three parameters (although the effect is now exponential).
Higher persistence and higher volatility reinforce each other. 
The impact of $\rho$ is nonlinear and large in the neighborhood of unity.


\section{Unbounded Rewards}

\label{s:ubr}

In this section we show that the optimality results presented above extend to
a range of unbounded reward settings after suitable modifications.
We consider the additively separable aggregator 
\begin{equation}
  \label{eq:H}
  H(x, z, x', v) = u(x, z, x') + \beta(z) \int v(x', z') Q(z, dz').
\end{equation}
The continuation value function $v$ is in $\vV$, which is the set of
all \emph{candidate value functions} and varies across applications. As
before, $\beta \in bc\ZZ$ and $Q$ is a Feller transition kernel.  The feasible
correspondence $\Gamma$ is assumed to be continuous, nonempty, and compact
valued. The reward function $u$ is continuous but not necessarily bounded.
The Euclidean norm is represented by $|\cdot|$.

\subsection{Homogeneous Functions}
\label{ss:hom}

We begin by extending the core results of \cite{alvarez1998dynamic} to the
case of state-dependent discounting.  We consider reward functions that are
homogeneous of degree $\theta \in (0, 1]$ and feasible correspondences that
are homogeneous of degree one.\footnote{Recall that a real-valued $f$ defined on a convex
cone $C$ of $\RR^k$ is homogeneous of degree $\theta$ if $f(\lambda x) =
\lambda^\theta f(x)$ for all $\lambda \geq 0$ and $x \in C$.} 

\begin{assumption}
  \label{a:hom1}
  $\XX$ is a convex cone in $\RR^k_+$ and 
  $\lambda x' \in \Gamma(\lambda x, z)$ when $(x, z, x')\in \graph \Gamma$
  and $\lambda \geq 0$. For each $z\in\ZZ$, $u(\cdot, z, \cdot)$ is
  homogeneous of degree $\theta$, and there exists a $B > 0$ such that 
  \begin{equation*}
    |u(x, z, x')| \leq B(|x| + |x'|)^\theta
    \text{ for all } (x, z, x')\in \graph \Gamma.
  \end{equation*}
\end{assumption}

Assumption~\ref{a:hom1} follows \cite{alvarez1998dynamic}.
The next assumption generalizes their growth restriction
to problems with state-dependent discounting.

\begin{assumption}
  \label{a:hom2}
  There exists an $\alpha \geq 0$ in $bm\ZZ$ such that $|x'|
  \leq \alpha(z) |x|$ when $(x, z, x')\in \graph \Gamma$. In
  addition, for $\{Z_t\}$ generated by $Q$,
  \begin{equation}
      \label{eq:ecc-hom}
      \sup_{z\in \ZZ} \, \EE_z \, \prod_{t=0}^{n-1}\beta(Z_t) \alpha^\theta(Z_t)< 1
      \; \text{ for some } n \in \NN.
  \end{equation}
\end{assumption}

The function $\alpha$ is a state-dependent upper bound on the growth rate of
the state variable. Comparing to the eventual discounting condition in
Section~\ref{ss:reg}, the extra term $\alpha^\theta(Z_t)$ in \eqref{eq:ecc-hom}
reflects the need to take into account the growth restriction when the
reward function is homogeneous and unbounded above. If both $\beta$ and
$\alpha$ are constant, then \eqref{eq:ecc-hom} reduces to the condition
$\alpha^\theta \beta < 1$ used in \cite{alvarez1998dynamic}.

In household problems where the state is asset holdings, the gross asset
return bounds the growth rate of the state. The condition
in~\eqref{eq:ecc-hom} implies that the shocks to the discount factor and
asset return have a similar effect on eventual discounting, but their
relative importance depends on the degree of homogeneity of the reward
function.


Let $(h_\theta \sS, \|\cdot\|_h)$ be the space of continuous functions on
$\sS$ that are homogeneous of degree $\theta$ in $x$ and bounded in the norm
defined by
\begin{equation}
  \label{eq:Hnorm}
  \|f\|_h := \sup
  \setntn{ |f(x, z)| }{ z\in \ZZ, \, x\in \XX, \, |x|=1 } .
\end{equation}
Then $h_\theta \sS$ is a Banach space \citep{stokey1989recursive}. To make
the problem well defined, we let $v_0 \equiv \textbf{0}$ so the
$\sigma$-value function is given by $v_\sigma := \lim_n
(T_\sigma\textbf{0})$. 

\begin{proposition}\label{p:hom}
  Let $\vV = h_\theta \sS$. Under Assumptions~\ref{a:hom1}--\ref{a:hom2}, the lifetime value $v_\sigma$ is well defined and
  finite on $\sS$ for any feasible policy $\sigma$,
  the value function $v^*$ is a unique fixed point of $T$ on $\vV$, $T^nv
  \to v^*$ for all $v\in \vV$, there exists an optimal policy
  that is homogeneous of degree one, and the principle of optimality holds.
\end{proposition}

\begin{example}\label{eg:1}
  Consider the household saving problem in \citet{toda2019wealth} where the
  exogenous state $\{Z_t\}$ is Markovian on $\ZZ$ with stochastic kernel $Q$. The
  asset return $R$ and discount function $\beta$ are bounded
  continuous functions of $Z_t$. The utility function is $u(c) =
  c^{1-\gamma}/(1-\gamma)$ with $\gamma \in (0, 1)$. The
  budget constraint is $X_{t+1} = R(Z_t)(X_t - C_t) \geq 0$ where $X_t$ is the
  beginning-of-period wealth and $C_t$ is consumption.  The Bellman equation is 
  \begin{equation*}
    v(x, z) = 
    \max_{c, x' \geq 0}
        \left\{ 
            u(c) + \beta(z)\int v\left(x', z'\right) Q(z, dz') 
        \right\} \; \st \; x' = R(z)(x-c).
  \end{equation*}
  If we use the constraint to eliminate $c$ in the Bellman equation
  and let $\Gamma(x, z) = [0, R(z)x]$, then
  Assumption~\ref{a:hom1} is satisfied with $\theta = 1-\gamma$ and $B =
  1/(1-\gamma)$. By Proposition~\ref{p:0}, Assumption~\ref{a:hom2} holds if
  $r(L_\alpha) < 1$ with $L_\alpha$ defined by
  \begin{equation*}
      (L_\alpha h)(z) := \beta(z) R^{1-\gamma}(z) \int h(z')Q(z, dz'),
  \end{equation*}
  where we let the upper bound function $\alpha = R$.
  This is a direct extension of the results in \cite{toda2019wealth} to the
  case of infinite $\ZZ$.  In particular, the condition $r(L_\alpha)<1$ reduces to 
  the condition in Proposition~1 of
  \citet{toda2019wealth} whenever $\ZZ$ is finite.
\end{example}


\subsection{Local Contractions}
\label{ss:loc}

Next we adopt a local contraction approach to dynamic programs with state
dependent discounting and unbounded rewards, extending methods first developed
in \cite{rincon2003existence}.  As in the previous section, the
aggregator has the form of \eqref{eq:H}. 

Let $c\sS$ be all continuous functions on $\sS$. Let $\ZZ$ be compact
and write $\XX = \bigcup_j \interior K_j$ where $\{K_j\}$ is a sequence of
strictly increasing and compact subsets of $\XX$. Let
\begin{equation*}
  \|f\|_j 
  := \sup_{x\in K_j, z\in \ZZ} |f(x, z)|
    \qquad (f\in c\sS).
\end{equation*}
Let $c>1$ and $\{m_j\}$ be an unbounded sequence of increasing positive real
numbers. Let $c_m\sS$ be all $f\in c\sS$ such that
\begin{equation*}
  \|f\|_m := \sum_{j=1}^\infty \frac{\|f\|_j}{m_j c^j} < \infty.
\end{equation*}
The pair $(c_m\sS, \|\cdot\|_m)$ forms a Banach space \citep{matkowski2011discounted}.

\begin{assumption}
  \label{a:loc2}
  $\Gamma(x, z) \subset K_j$ for all $x\in K_j$, all $z \in \ZZ$, and all
  $j\in \NN$, and $(\beta, Q)$ is eventually discounting in the sense of
  Section~\ref{ss:reg}.
\end{assumption}

\begin{proposition}
  \label{p:loc}
  Under Assumption~\ref{a:loc2}, the lifetime value
  $v_\sigma$ is well defined and finite on $\sS$ for any 
    $\sigma \in \Sigma$, there exists a 
  sequence $m_j \uparrow \infty$ such that the value function $v^*$ is the
  unique fixed point of $T$ on $c_m\sS$, $T^nv \to v^*$ for all $v\in
  c_m\sS$, there exists an optimal policy, and the principle of
  optimality holds.
\end{proposition}

\begin{example}
    Consider a stochastic optimal growth model with state dependent
    discounting, total production $zf(x)$ and continuous utility $u$.
    The feasible correspondence is $\Gamma(x, z) = [0, zf(x)]$.
    Let $\XX = \RR_+$ and let $\ZZ \subset \RR_+$ be compact.
    Suppose $f' > 0$, $f'' < 0$ and $\lim_{x\to \infty}f'(x)
    = 0$. Let $\{K_j\}$ be an increasing sequence of 
    compact sets covering $\XX$ such that $\Gamma(x, z) \subset K_j$ for all
    $x\in K_j$.\footnote{For example, set $K_j := [0, M+j]$ for all $j \in
      \NN$, where $M$ is some large constant.} 
      Assumption~\ref{a:loc2} holds and
      Proposition~\ref{p:loc} can be applied if $(\beta, Q)$ is eventual discounting.
\end{example}

\section{Further Extensions}\label{s:ext}

We study two further extensions. Section~\ref{ss:alt}
studies an alternative discount specification to the
framework in Section~\ref{s:model}. Section~\ref{ss:ez} extends our main
results to Epstein-Zin preferences with unbounded rewards.

\subsection{Alternative Discount Specifications}\label{ss:alt}

Discounting methods that differ from the preceding framework can also be analyzed.
To illustrate, we consider 
the shocks to long-run discount factors found in 
 \cite{primiceri2006intertemporal}, \cite{justiniano2010investment},
\cite{leeper2010government}, and \cite{christiano2014risk}.
Their maximization problems are analogous to
the additively separable problem in Section~\ref{ss:genadsep}, with the
difference that $\prod_{t=0}^{n-1} \beta_t$ is replaced by
$b^n Z_n$ for some constant $b$. While the discount factor $b^n Z_n$ can be expressed as
$\prod_{t=0}^{n-1} \beta_t$ after setting $\beta_t := b Z_{t+1}/Z_t$ and $Z_0
= 1$, notice that $\beta_t$ is not observable until $t+1$.  
Hence inequality \eqref{eq:bc} cannot be used, since it 
assumes that $\beta_t$ is visible at $t$.

To handle such cases, one option is to replace inequality~\eqref{eq:bc} with 
\begin{equation}\label{eq:bc2}
    |H(x, z, x', v) - H(x, z, x', w)| 
    \leq \int \beta(z') |v(x', z') - w(x', z')|Q(z, dz').
\end{equation}
Inequality~\eqref{eq:bc2} integrates over $\beta(z')$, supposing that its
realization is not observed at the time that $x'$ is chosen.  We prove in the
appendix that Theorem~\ref{t:bk} extends to this case: the theorem is valid
under eventual discounting when~\eqref{eq:bc2} replaces \eqref{eq:bc}.

The set up of \cite{primiceri2006intertemporal} and other authors mentioned
above satisfies~\eqref{eq:bc2} after redefining the
aggregator and the exogenous state variable.\footnote{\label{fn:redef}To be
  specific, let the exogenous state variable be $\tilde{Z}_{t+1} = (Z_{t+1},
  Z_t)$. The aggregator then becomes $H(x, z, x', v) = \tilde{F}(x, z, x') +
  \int \beta(z') v(x', z') \tilde{Q}(z, dz')$, where $\tilde{F}(X_t,
  \tilde{Z}_t, X_{t+1}) = F(X_t, Z_t, X_{t+1})$, $\beta(\tilde{Z}_{t+1}) =
  bZ_{t+1}/Z_t$, and $\tilde{Q}$ is the transition kernel on $\tilde{\ZZ} :=
  \ZZ^2$ induced by $Q$.} The only question, then, is whether or not eventual
  discounting holds.  The following proposition shows that, in many cases, 
  the answer depends only on the value of $b$ in $\beta_t := b Z_{t+1}/Z_t$.
  Stochastic components are irrelevant.

\begin{proposition}\label{p:shocks}
    If $\beta_t := b Z_{t+1}/Z_t$ for all $t$ and $\{Z_t\}$ is positive and
    bounded, then eventual discounting holds if and only if $b<1$.
\end{proposition}

The intuition behind Proposition~\ref{p:shocks} is that
the spectral radius $r(L_\beta)$ equals the asymptotic growth rate of the discount factor
process.  If $\prod_{t=0}^{n-1} \beta_t = b^n Z_n$ and $Z_t$ is positive and bounded, the
asymptotic growth rate is equal to $b$.


\subsection{Epstein-Zin Preferences}
\label{ss:ez}

Next we extend the preceding results on dynamic programming 
under state-dependent discounting to settings where lifetime utility is
governed by Epstein--Zin preferences.  Lifetime utility of an agent satisfies
\begin{equation}
  \label{eq:koopmans}
  U(C_t, C_{t+1}, \ldots) = \left\{ C_t^{1-1/\psi} + \beta_t \left[\EE_t
      U^{1-\gamma}(C_{t+1}, C_{t+2},
      \ldots)\right]^{\frac{1-1/\psi}{1-\gamma}} \right\}^{\frac{1}{1-1/\psi}},
\end{equation}
where $\gamma$ is the relative risk aversion and $\psi$ is the elasticity of
intertemporal substitution. The agent maximizes lifetime utility by choosing
consumption $\{C_t\}$ subject to $X_{t+1} = R_t (X_t -
C_t) \geq 0$. Here $X_t$ is asset holding of the agent at the beginning of
time $t$ and $R_t$ is returns.
We focus on the empirically relevant case of $\gamma > 1$ and $\psi>1$, as
in, say, \cite{bansal2004risks}, \cite{albuquerque2016valuation}, or
\cite{schorfheide2018identifying}. This is the most challenging setting
because the usual contraction argument fails and the utility function is
unbounded above.

\subsubsection{Discounting Continuation Values}

Let $\XX = \RR_+$, assume that
$\beta_t$ and $R_t$ are functions of the exogenous state, and define the
aggregator $H$ by
\begin{equation}
  \label{eq:EZ-H}
  H(x, z, c, v) = \left\{c^{1-1/\psi} + \beta(z) \left[ \int v\left(R(z)(x-c),
        z'\right)^{1-\gamma} Q(z, dz') \right]^{\frac{1-1/\psi}{1-\gamma}}
  \right\}^{\frac{1}{1-1/\psi}},
\end{equation}
where $x$, $z$, and $c$ are asset holding, exogenous state, and consumption,
respectively, satisfying $c \in \Gamma(x, z) = [0, x]$.

\begin{assumption} \label{a:EZ}
    The functions $\beta$ and $R$ are nonnegative elements of $bm\ZZ$. In
    addition, for $\{Z_t\}$ generated by $Q$, we have
    \begin{equation}
        \label{eq:ecc-EZ}
        \sup_{z\in \ZZ} \, \EE_z \,
        \prod_{t=0}^{n-1}\beta(Z_t)^{1/(1-1/\psi)} R(Z_t)< 1
        \; \text{ for some } n \in \NN.
    \end{equation}
\end{assumption}

Assumption~\ref{a:EZ} is an eventual discounting condition for the Epstein--Zin case.
It is modified to take into account both the underlying growth rate, as in
Assumption~\ref{a:hom2}, and also the role of elasticity of intertemporal
substitution.  (Intuition and numerical applications are provided below.)

Let $\vV$ be all $f \in m\sS$ such that
$\|f\|_I := \sup_{x\in\XX, z\in\ZZ}|f(x, z)/(1+x)|$ is finite.
We show in Appendix~\ref{ss:EZ-proofs} that there exists an upper bound function $\hat{v}
\in \vV$ such that $T_\sigma$ is a self map on the order interval $[0,
\hat{v}] \subset \vV$ with the pointwise partial order. Then we show that
$v_\sigma$ is well defined on the order interval and is a fixed point of
$T_\sigma$. In addition, if $\sigma$ satisfies an interiority condition, the
fixed point is unique. See Proposition~\ref{p:v_sigma}.

Let $\hat{\vV}$ be the space of functions in $\vV$ that are homogeneous of degree one in $x$. 
Our main result for this section is as follows.

\begin{proposition}
  \label{p:vstar}
  If Assumption~\ref{a:EZ} holds, then $\bar{v} :=
  \lim_{n \to \infty} T^n \mathbf{0}$ is a well defined element of $\hat{\vV}$ and equal to the value function.
  There exists an optimal policy $\sigma^*\in\Sigma$ that is homogeneous of
  degree one in $x$ and the principle of optimality holds.
\end{proposition}

Notice that Proposition~\ref{p:vstar} contains no analogue of the eventual
contraction condition in Assumption~\ref{a:con}.  This is because, as
mentioned above, $T$ and $T_\sigma$ are not contraction mappings under
conventional metrics.  Instead, the proof uses monotonicity and a form of
concavity inherent in Epstein--Zin preferences, combined with fixed point results due to
\cite{marinacci2010unique}.

\subsubsection{Alternative Preference Shocks}\label{sss:aps}

While \eqref{eq:koopmans} parallels the definitions in, say,
\cite{epstein1989substitution}, \cite{nakata2020equilibrium} and \cite{de2020valuation}, other studies
introduce preference shocks to current consumption \citep{albuquerque2016valuation, schorfheide2018identifying}. In this setting, lifetime
utility satisfies
\begin{equation}
  \label{eq:koopmans2}
  U(C_t, C_{t+1}, \ldots) = \left\{ \lambda_t C_t^{1-1/\psi} + b
    \left[\EE_t U^{1-\gamma}(C_{t+1}, C_{t+2},
      \ldots)\right]^{\frac{1-1/\psi}{1-\gamma}} \right\}^{\frac{1}{1-1/\psi}},
\end{equation}
where $b < 1$ is a fixed constant and $\{\lambda_t\}$ is a preference
shock.\footnote{Some authors also place an additional term $(1-b)$
  before $\lambda_t$. This is inconsequential to our optimality results
  since we can simply redefine $\lambda_t$ to include $(1-b)$.}
As we now show, the preceding analysis can be brought to bear on this case as
well.



Using homogeneity and dividing both
sides of \eqref{eq:koopmans2} by $\lambda_t^{1/(1-1/\psi)}$ yields
\begin{equation}
  \label{eq:koopmans-alt}
  \tilde{U}_t = \left\{ C_t^{1-1/\psi} + b \left[\EE_t
      \tilde{U}_{t+1}^{1-\gamma}
      \left(\frac{\lambda_{t+1}}{\lambda_t}\right)^{\frac{1-\gamma}{1-1/\psi}}
    \right]^{\frac{1-1/\psi}{1-\gamma}} \right\}^{\frac{1}{1-1/\psi}}
\end{equation}
where $\tilde{U}_t := U(C_t, C_{t+1}, \ldots)/\lambda_t^{1/(1-1/\psi)}$.
If $\lambda_{t+1}/\lambda_t$ is measurable with respect to the time-$t$
information set, then~\eqref{eq:koopmans-alt} becomes
\begin{equation}\label{eq:tr} 
  \tilde{U}_t = \left\{C_t^{1-1/\psi} + b\delta_t \left[\EE_t
      \tilde{U}_{t+1}^{1-\gamma} \right]^{\frac{1-1/\psi}{1-\gamma}}
  \right\}^{\frac{1}{1-1/\psi}},
\end{equation}
where $\delta_t := \lambda_{t+1}/\lambda_t$. This is the same as the
original Koopmans equation in \eqref{eq:koopmans} with $\beta_t = b
\delta_t$.\footnote{The equivalence between $\beta_t$ and $b
  \lambda_{t+1}/\lambda_t$ is demonstrated in \cite{de2020valuation} using
  the Euler equation in an expected utility setting.}  Optimality
  results from the previous section can now be applied.

\subsubsection{Interpretation}\label{sss:int}

Condition~\eqref{eq:ecc-EZ} is the key restriction
required for Proposition~\ref{p:vstar} and elasticity of intertemporal substitution plays a role.  To illustrate the implications of
the condition we consider the study of \cite{albuquerque2016valuation}, who
adopt the specification in \eqref{eq:koopmans2} with $\delta_t :=
\lambda_{t+1}/\lambda_t$ satisfying $\log \delta_t = \rho \log \delta_{t-1} +
\sigma_\epsilon \epsilon_t$. 
In view of the discussion in
Section~\ref{sss:aps}, we can study optimality by applying the eventual discounting 
condition~\eqref{eq:ecc-EZ} to the transformed representation~\eqref{eq:tr}.
By a result analogous to Proposition~\ref{p:0}, condition~\eqref{eq:ecc-EZ} is
equivalent to $r(L_R) < 1$ with $L_R$ defined
by
\begin{equation}\label{eq:lrdef}
    (L_R h)(z) := \beta(z)^{1/(1-1/\psi)} R(z) \int h(z')Q(z, dz').
\end{equation}
One way to obtain insight on the value $r(L_R)$ is to use the stationary
approximation $s := \lim_{n\to\infty} s_n^{1/n}$, where $s_n := \EE \prod_{t=0}^{n-1} \beta_t^{1/(1-1/\psi)} R_t$.
The advantage of the stationary approximation is that, if we specialize to 
$R(z) \equiv R$, then we obtain the analytical expression
\begin{equation}\label{eq:EZ-r}
    s = R \exp \left(
      \frac{1}{1-1/\psi} \log b + \frac{1}{(1-1/\psi)^2}
      \frac{\sigma_\epsilon^2}{2(1-\rho)^2}
    \right).
\end{equation}

(See Appendix~\ref{s:geomean} for details.) Analogous to the findings in
Section~\ref{sss:ar1logs} (cf.~Table~\ref{tab:r}), this stationary
representation closely approximates $r(L_R)$ for a discretized version with
moderately fine grid.

The expression in~\eqref{eq:EZ-r} sheds light on the role that elasticity of intertemporal substitution
plays in eventual discounting. 
The impact of $\psi$ in \eqref{eq:EZ-r} is not monotone because the mean
term $\log b$ is typically negative, while the volatility term $\sigma_\epsilon^2 /
(2(1-\rho)^2)$ is positive. Nonetheless, we can understand the impact of $\psi$ by the
relative weight placed on the mean and volatility terms: $1/(1-1/\psi)$
enters \eqref{eq:EZ-r} directly for the mean and is squared on the
volatility term. Hence, as $\psi$ rises and $1/(1-1/\psi)$ falls, the
relative importance of $b$ in determining $r(L_R)$ increases. 
Conversely, as $\psi \downarrow 1$, the volatility term increasingly dominates.

Intuitively, if
 $\psi$ is large, then the agent is more willing
to shift consumption across time, so the volatility in the
discount factor plays a lesser role. Conversely, when $\psi$ is small, 
 consumption cannot shift as freely to compensate for
fluctuations in the discount factor. Hence volatility in the discount factor
has a large impact on lifetime utility.

\subsubsection{Numerical Analysis}\label{sss:mc}

In the applications discussed in Section~\ref{ss:finite}, discount dynamics
are driven by Gaussian AR(1) processes, where standard discretization methods are available and 
eventual discounting is easy to test.  In some recent studies,
however, discounting is driven by a Markov process and
additional innovations, as in \cite{albuquerque2016valuation}, or 
stochastic volatility, as in \cite{basu2017uncertainty}. For such cases, one
can either use a more sophisticated discretization procedure (see, e.g.,
\cite{farmer2017discretizing})  or use Monte Carlo.

To illustrate the Monte Carlo method, we return to the model in
\cite{albuquerque2016valuation} studied above,
where the eventual discounting condition is
\eqref{eq:ecc-EZ}, or equivalently, $r(L_R) < 1$ with $L_R$ defined in
\eqref{eq:lrdef}. An analytical expression was obtained in
\eqref{eq:EZ-r} for the case when $R_t$ is constant, but in
\cite{albuquerque2016valuation} this is not the case.  Nonetheless,
by the strong law of large numbers, we can
approximate each $s_n$ by generating $m$ independent simulated paths of
$\{\beta_t, R_t\}$ and calculating
\begin{equation}\label{eq:mc_alb}
    \hat{s}_n = \frac{1}{m} \sum_{i=1}^m \prod_{t=0}^{n-1}
    \beta_{i, t}^{1/(1-1/\psi)}R_{i, t}.
\end{equation}
Using the parameters in \cite{albuquerque2016valuation}, we find that
$\hat{s}_n^{1/n}$ increases with $n$ and exceeds one when $n$ is large, as
shown in Table~\ref{tab:mc}.\footnote{We treat the baseline model in
  \cite{albuquerque2016valuation}, where $\gamma = 1.516$ and $\psi =
  1.4567$. There are three exogenous states: preference shock $x_t$, log
  consumption growth $\Delta c_{t}$, and log price consumption ratio
  $z_{ct}$. The discount factor is $\beta_t = b e^{x_t}$ with $x_t = \rho
  x_{t-1} + \sigma \epsilon_t$, $b=0.99795$, $\rho = 0.99132$, and $\sigma =
  0.00058631$. The logarithm of returns satisfies $r_{t+1} = \kappa_{c0} +
  \kappa_{c1} z_{ct+1} - z_{ct} + \Delta c_{t+1}$ where $z_{ct} = A_{c0} +
  A_{c1} x_t$ and $\Delta c_{t+1} = \mu + \sigma_c \epsilon^c_{t+1}$ with
  $\mu = 0.0015644$ and $\sigma_c = 0.0069004$. The remaining parameters can
  be solved as detailed in their Internet Appendix, giving $\kappa_{c0} =
  0.023108, \kappa_{c1} = 0.99653, A_{c0} = 5.6605$, and $A_{c1} = 82.519$.
  We run a large number of simulations ($m=100000$) for each experiment to
  ensure that $\hat{s}_n$ is close to $s_n$. The last row lists the standard
  error for each estimate by calculating the standard deviation of 1000
  simulated $\hat{s}_n^{1/n}$ with $\hat{s}_n$ replaced by an approximating
  normal distribution for computational efficiency.} This is in line with
the analytical expression given by~\eqref{eq:EZ-r}, which yields $s =
1.0168$ if we fix $R_t \equiv 1$. Hence eventual discounting fails under their
parameterization.\footnote{We have not shown the eventual discounting condition to be necessary in the
Epstein--Zin case, so the optimization problem in
\cite{albuquerque2016valuation} might still be well defined.  The quantitative
exercise in \cite{albuquerque2016valuation} does not shed light on this issue
because they do not solve the agent's optimization problem
directly. Instead, they assume that a solution exists and use it to derive
asset pricing moments.}

\begin{table}[tb!]
    \caption{Calculate $r(L_R)$ Using Monte Carlo Method}
    \label{tab:mc}
    \footnotesize
    \begin{tabular*}{\textwidth}{@{} l @{\extracolsep\fill} c @{\extracolsep\fill} c @{\extracolsep\fill} c @{\extracolsep\fill} c @{}}
      \toprule
      Length of Paths & $n=100$   & $n=200$   & $n=500$   & $n=1000$  \\
      \midrule
      Estimate of $r(L_R)$        & 1.00355   & 1.00698   & 1.01220   & 1.01321   \\
      Standard Error  & (0.00004) & (0.00008) & (0.00045) & (0.00054) \\
      \bottomrule
    \end{tabular*}
\end{table}

\subsubsection{The Role of Elasticity of Intertemporal Substitution}\label{sss:ies}

In a New Keynesian model with preference similar to \eqref{eq:koopmans2}
studied by \cite{basu2017uncertainty},
\cite{de2018uncertainty} show that the responses to discount factor
shocks explode when the elasticity of intertemporal substitution approaches
one, and that this issue disappears if $\beta_t$ is constant. This matches
\eqref{eq:EZ-r}. If the volatility term is not zero, then $r(L_R)$ becomes arbitrarily
large as $\psi$ approaches one.  Hence it appears that the
large responses found in \cite{de2018uncertainty} are the result
of an ill-defined household problem that fails the eventual discounting
condition. If $\beta_t \equiv b$, then \eqref{eq:EZ-r} becomes
$b^{1/(1-\psi)} R$. Letting $\psi$ approach one will push down $r(L_R)$
instead so the issue disappears.

In \cite{de2018uncertainty}, the asymptote in the responses is attributed to the
distributional weights on current and future utility not summing to one.
They propose an alternative setting where current utility is weighted by
$1-\beta_t$ and future utility is weighted by $\beta_t$ with $\beta_t < 1$.
We show in the appendix that the eventual discounting condition for this
specification is the same as Assumption~\ref{a:EZ}. Since $\beta_t$ is
assumed to be strictly less than one in \cite{de2018uncertainty}, we let
$\beta_t \leq b$ for
some $b < 1$ and assume fixed returns. Then \eqref{eq:EZ-r} implies that
$r(L_R) \leq b^{1/(1-1/\psi)} R$. The previous discussion shows that, in this
case, eventual discounting holds when $\psi$ approaches one. This provides
an alternative explanation of why the model does not
produce an asymptote in responses to discount factor shocks.

\section{Conclusion}\label{s:c}

We introduce a weak discounting condition and show that, under this condition, standard infinite horizon dynamic
programs with state-dependent discount rates are well defined and well
behaved.   The value function
satisfies the Bellman equation, an optimal policy exists, Bellman's principle
of optimality is valid, value function iteration converges and so does
Howard's policy iteration algorithm.  The method can be applied to a broad
range of dynamic programming problems, including those with discrete choices,
continuous choices and recursive preferences.

We connect eventual discounting to a spectral radius condition and
provided guidelines on how to calculate the spectral radius for a range of
discount specifications. We show that the condition is more likely to fail
when the discount process has higher mean, persistence, or volatility.
For models with Epstein--Zin preferences and state-dependent discount factors,
the condition also depends on the elasticity of intertemporal substitution.


One natural open question is: how do our results translate into continuous
time?    It would also be
valuable to understand how the results change if discounting depends on
endogenous states and actions.  Finally, more research is needed on how close
to necessary the eventual discounting conditions are for recursive preference
models, and especially those involving long run risks, since these models
generate realistic asset price processes by driving their parameterizations
close to the boundary  between stability and instability.  These questions are
left to future research.

\appendix
\section{Remaining Proofs}

In what follows, we consider the dynamic program described in
Section~\ref{ss:fr}.

\subsection{Proofs for Section~\ref{s:model}}
\label{s:re-proofs}

\subsubsection{Proof of Theorem~\ref{t:bk}}

For each $\sigma\in\Sigma$, let $T_\sigma$ be defined on $bm\sS$ by
\eqref{eq:T_sigma}. Let $T$ be defined on $bc\sS$ by \eqref{eq:bellman}. We
prove part~\ref{bk-a} through two lemmas.

\begin{lemma}
  \label{l:T_sigma}
  If $\sigma \in \Sigma$, then $T_\sigma$ is eventually contracting on
  $bm\sS$.
\end{lemma}

\begin{proof}
    Fix $\sigma \in \Sigma$ and $v \in bm\sS$.  The map $T_\sigma v$ is Borel measurable on $\sS$
    by the regularity conditions and measurability of $\sigma$.  It is bounded
    by the assumption that $H$ is bounded.  Hence $T_\sigma$ is a self-map on
    $bm\sS$.  To see that it is eventually contracting, 
    fix $(x, z)$ in $\sS$ and observe that, by 
    Assumption~\ref{a:con}, 
    \begin{align*}
      |(T_\sigma v)(x, z) - (T_\sigma w)(x, z)|
        & = \left| H(x, z, \sigma(x, z), v) - H (x, z, \sigma(x, z), w) \right|\\
        & \leq \beta(z)\int |v(\sigma(x, z), z')
        - w(\sigma(x, z), z')|Q(z, dz')
    \end{align*}
    for any $v, w\in bm\sS$.
    We can write this expression as 
    \begin{equation}\label{eq:tbk}
      |T_\sigma v  - T_\sigma w| \leq K_\sigma | v  -  w|,
    \end{equation}
    where $K_\sigma$ is the operator defined by
    \begin{equation*}
      (K_\sigma h)(x, z) := \beta(z) \int h(\sigma(x, z), z') Q(z, d z')
      \qquad (h \in bm\sS, z \in \ZZ).
    \end{equation*}
    Since $\beta\in bc\ZZ$, $K_\sigma$ is a self-map on $bm\sS$. Since
    $K_\sigma$ is order preserving, we can iterate on~\eqref{eq:tbk} to
    obtain $|T_\sigma^n v - T_\sigma^n w| \leq K_\sigma^n | v - w|$ for all
    $n \in \NN$.


    Let $\{Z_t\}$ be a Markov process generated by $Q$ and started at $z$,
    let $\beta_t = \beta(Z_t)$, and let $\{X_t\}$ be the controlled Markov
    process generated by $X_{t+1} = \sigma(X_t, Z_t)$ with $(X_0, Z_0) = (x,
    z)$. We then have $(K_\sigma h)(x, z) = \EE_{x, z} \, \beta_0 \, h(X_1,
    Z_1)$ and, iterating on this equation,
    \begin{equation}\label{eq:knb}
        (K_\sigma^n h)(x, z) 
        = \EE_{x, z} \, \beta_0 \beta_1 \cdots \beta_{n-1} \, h(X_n, Z_n)
        \leq r_n^\beta \| h \|.
    \end{equation}
    Since $|T_\sigma^n v - T_\sigma^n w| \leq K_\sigma^n | v - w|$, taking
    the supremum yields $\|T_\sigma^n v - T_\sigma^n w\| \leq r_n^\beta \|v-w\|$.
    It now follows from the eventual discounting property that $T_\sigma^n$ is a contraction
    for some $n \in \NN$. Hence $T$ is eventually contracting.
\end{proof}

\begin{lemma}
  \label{l:T}
  The operator $T$ is eventually contracting on $bc\sS$.
\end{lemma}

\begin{proof}
  Fix $v \in bc\sS$. The map $T v$ is continuous on $\sS$ by regularity and
  Berge's Maximum Theorem \citep[Theorem~17.31]{aliprantis2006infinite}. It
  is bounded by boundedness of $H$. Hence $T$ is a self-map on $bc\sS$. To
  see that it is eventually contracting, fix $(x, z)$ in $\sS$ and observe
  that, by Assumption~\ref{a:con},
    \begin{align*}
      |(T v)(x, z) - (T w)(x, z)|
        & \leq \max_{x' \in \Gamma(x, z)} 
            \left|H(x, z, x', v) - H(x, z, x', w) \right|
            \\
        & \leq \max_{x' \in \Gamma(x, z)} \beta(z)
            \int  |v(x', z') - w(x', z')| \, Q(z, dz')
    \end{align*}
    for any $v, w\in bc\sS$. We can write this expression as
    \begin{equation}\label{eq:bbm}
      |T v  - T w| \leq K | v  -  w|,
    \end{equation}
    where $K$ is the operator on $bc\sS$ defined by
    \begin{equation*}
        (Kh)(x, z) := \max_{x' \in \Gamma(x, z)}\beta(z) \int h(x', z')
        Q(z, d z')
        \qquad (h \in bc\sS, z \in \ZZ).
    \end{equation*}
    It follows from regularity and the Feller property (see footnote \ref{fn:feller})
    that $(x', z) \mapsto \int h(x', z')Q(z, dz')$ is continuous. Since
    $\beta\in bc\ZZ$, it follows from the maximum theorem that $K$ is a
    self-map on $bc\sS$. Since $K$ is order preserving, we can iterate
    on~\eqref{eq:bbm} to obtain $|T^n v - T^n w| \leq K^n | v - w|$ for all
    $n \in \NN$.

    Now set $h := |v - w|$, let $\{Z_t\}$ be a Markov process generated by
    $Q$ with initial condition $z$ and let $\beta_t = \beta(Z_t)$. We then
    have $(Kh)(x, z) = \max_{x_1 \in \Gamma(x, z)} \EE_z \, \beta_0 \,
    h(x_1, Z_1)$ and hence
    \begin{align*}
      (K^2 h)(x, z) & = \max_{x_1 \in \Gamma(x, z)} \EE_z \, \beta_0 \,
      (Kh)(x_1, Z_1)
      \\
      & = \max_{x_1 \in \Gamma(x, z)} \EE_z \, \beta_0 \, \max_{x_2 \in
        \Gamma(x_1, Z_1)} \EE_{Z_1} \, \beta_1 \, h(x_2, Z_2)
      \\
      & \leq \| h \| \, \EE_z \, \beta_0 \beta_1.
    \end{align*}
    More generally, for arbitrary $n \in \NN$, we have $(K^n h)(x, z) \leq
    r^\beta_n
    \| h \|$.
    Since $|T^n v  - T^n w| \leq K^n h$, taking the supremum gives $\|T^n v  -
    T^n w\| \leq r^\beta_n \| v - w \|$ for all $n \in \NN$.  It follows from
    eventual discounting that $T^n$ is a contraction for some $n\in\NN$ and hence
    $T$ is eventually contracting.
\end{proof}

We have an immediate corollary to Lemma~\ref{l:T_sigma} and \ref{l:T}.

\begin{corollary}
  \label{c:fix}
  If $v_0\in bm\sS$, the $\sigma$-value function $v_\sigma$ is the unique
  fixed point of $T_\sigma$ in $bm\sS$ and $T_\sigma^n v \to v_\sigma$ for
  all $v\in bm\sS$. The Bellman operator $T$ has a unique fixed point
  $\bar{v}$ in $bc\sS$ and $T^n w \to \bar{v}$ for all $w\in bc\sS$.
\end{corollary}

\begin{proof}
  By Lemma~\ref{l:T_sigma} and a generalized Contraction Mapping Theorem
  \citep[see, e.g.,][Section 4.2]{cheney2013analysis}, $T_\sigma$ is
  globally stable on $bm\sS$. Hence, if $v_0\in bm\sS$, $v_\sigma$ is the
  unique fixed point of $T_\sigma$ in $bm\sS$ and $T_\sigma^n v \to
  v_\sigma$ for all $v\in bm\sS$. The claim for $T$ follows similarly from
  Lemma~\ref{l:T}.
\end{proof}

Part~\ref{bk-b} follows directly from Corollary~\ref{c:fix}.

Next we show that $\bar{v}$ given by Corollary~\ref{c:fix} is the value
function. First note that $\bar{v} = T\bar{v} \geq T_\sigma \bar{v}$ by
definition. Iterating $T_\sigma$ on both sides and using
\eqref{eq:mon}, we have $\bar{v} \geq T_\sigma^n \bar{v}$. Taking
$n$ to infinity, it follows from Corollary~\ref{c:fix} that $\bar{v} \geq
v_\sigma$. Taking the supremum over $\Sigma$ gives $\bar{v} \geq v^*$.

For the other direction, regularity and the measurable maximum theorem
\citep[Theorem 18.19]{aliprantis2006infinite} ensure that there exists a
$\sigma^* \in \Sigma$ such that $T_{\sigma^*} \bar{v} = T\bar{v}$. Then we
have $T_{\sigma^*} \bar{v} = \bar{v}$. Because $\bar{v}\in bc\sS \subset
bm\sS$ and $T_{\sigma^*}$ has a unique fixed point in $bm\sS$ by
Corollary~\ref{c:fix}, $\bar{v} = v_{\sigma^*}$. By the definition of $v^*$,
we have $v^* \geq v_{\sigma^*} = \bar{v}$. Therefore, $v^* = \bar{v}$ and
$\sigma^*$ is the optimal policy. This proves \ref{bk-c} and \ref{bk-d}.

One direction of the Bellman's principle of optimality is implied in the
argument above. For the other direction, if a policy $\sigma$ is optimal,
then $v_\sigma = v^*$. It follows from Corollary~\ref{c:fix} that $v^* =
T_\sigma v^*$. Since $v^* = \bar{v}$ is the fixed point of $T$, $T_\sigma
v^* = T v^*$. This proves \ref{bk-e}.

Part \ref{bk-f} is valid by Corollary~\ref{c:fix} and the fact that $\bar{v}
= v^*$.

For part \ref{bk-g}, the following proof is adapted from \citet[Proposition
2.4.1]{bertsekas2013abstract}.

Suppose there exists $\{\sigma_k\}\subset\Sigma$ such that $T_{\sigma_k}
v_{\sigma_{k-1}} = T v_{\sigma_{k-1}}$. By definition, $T_{\sigma_k}
v_{\sigma_{k-1}} = T v_{\sigma_{k-1}} \geq T_{\sigma_{k-1}} v_{\sigma_{k-1}}
= v_{\sigma_{k-1}}$. By inequality~\eqref{eq:mon}, applying $T_{\sigma_k}$ to
both sides repeatedly gives $T^n_{\sigma_k} v_{\sigma_{k-1}} \geq T
v_{\sigma_{k-1}} \geq v_{\sigma_{k-1}}$. Taking $n$ to infinity, it follows
from Corollary~\ref{c:fix} that $v_{\sigma_k} \geq T v_{\sigma_{k-1}} \geq
v_{\sigma_{k-1}}$. An inductive argument implies that $v ^* \geq
v_{\sigma_k} \geq T^k v_{\sigma_0}$. Taking $k$ to infinity,
Corollary~\ref{c:fix} then implies that $v_{\sigma_k} \to v^*$.

\subsubsection{Blackwell's Condition}

\begin{proposition}[Blackwell's Condition]
  \label{p:blackwell}
  Let $\dD = (\Gamma, H)$ be a regular dynamic program. If there exists a
  nonnegative function $\beta \in bc\ZZ$ and a Feller transition kernel $Q$
  on $\ZZ$ such that $(\beta, Q)$ is eventually discounting and
  the Bellman operator satisfies
  \begin{equation}
    \label{eq:blackwell}
    [T(v+c)](x, z) \leq (Tv)(x, z) + \beta(z)\int c(z')Q(z, dz')
  \end{equation}
  for all $(x, z) \in \sS$, $v\in bc\sS$, and $c \in bm\ZZ_+$, then $T$ is
  eventually contracting on $bc\sS$.
\end{proposition}

\begin{proof}[Proof of Proposition~\ref{p:blackwell}]
  For any $v, w\in bc\sS$, we have
  \begin{equation*}
    v(x, z) - w(x, z) \leq \sup_{x'\in\XX}|v(x', z) - w(x', z)| =: c(z)
  \end{equation*}
  for all $(x, z)\in\sS$, where $c$ is lower semicontinuous \citep[Lemma
  17.29]{aliprantis2006infinite} and thus $c \in bm\ZZ_+$.
  Inequality~\eqref{eq:blackwell} implies that
  \begin{equation*}
    [T(v+c)](x, z) \leq (Tv)(x, z) + \beta(z)\int \sup_{x'\in\XX}|v(x', z')
    - w(x', z')|Q(z, dz').
  \end{equation*}
  It then follows from \eqref{eq:mon} that
  \begin{equation*}
    (Tv)(x, z) \leq (Tw)(x, z) + \beta(z) \int \sup_{x'\in\XX}|v(x', z') -
    w(x', z')|Q(z, dz').
  \end{equation*}
  Exchanging the roles of $v$ and $w$, we have
  \begin{equation*}
    |(Tv)(x, z) - (Tw)(x, z)| \leq \beta(z) \int \sup_{x'\in\XX}|v(x', z') -
    w(x', z')|Q(z, dz').
  \end{equation*}
  Iterating on the above inequality, it follows from a similar argument to
  the proof of Lemma~\ref{l:T} that
    $\|T^n v - T^n w\| \leq r^\beta_n \|v-w\|$.  Since $T$ is a self map
  on $bc\sS$, it follows from eventual discounting that $T$ is eventually
  contracting.
\end{proof}

\subsubsection{Monotonicity, Concavity, and Differentiability}

\begin{proof}[Proof of Theorem~\ref{t:mon}]
  Since $ibc\sS$ is a closed subset of $bc\sS$, it suffices to show that $T$
  maps $ibc\sS$ to functions in $ibc\sS$ that are strictly concave in $x$.
  For monotonicity, pick any $z\in\ZZ$ and $v\in ibc\sS$. Then for any $y
  \geq x$,
  \begin{align*}
    (Tv)(y, z) &= H\left(y, z, \sigma^*(y, z), v\right)\\
    &\geq H\left(y, z, \sigma^*(x, z), v\right)\\
    &\geq H\left(x, z, \sigma^*(x, z), v\right) = (Tv)(x, z),
  \end{align*}
  where the first inequality holds because $\sigma^*(x, z) \in \Gamma(x, z)
  \subset \Gamma(y, z)$ and the second inequality holds because $H$ is
  increasing in $x$ by Assumption~\ref{a:ibc}. For concavity, pick any
  $x, y$ satisfying $x\neq y$ and $\theta\in (0, 1)$ and define $x_\theta
  = \theta x + (1-\theta)y$. Then, for any $z\in\ZZ$ and $v\in ibc\sS$,
  \begin{align*}
    \theta (Tv)(x, z) + (1-\theta) (Tv)(y, z) &= \theta H\left(x, z,
      \sigma^*(x, z), v\right) + (1-\theta)H\left(y, z, \sigma^*(y, z),
      v\right)\\
    &< H\left(x_\theta, z, \theta\sigma^*(x, z) + (1-\theta) \sigma^*(y,
      z), v\right)\\
    &\leq H\left(x_\theta, z, \sigma^*(x_\theta, z), v\right) =
    (Tv)(x_\theta, z),
  \end{align*}
  where the first inequality holds because $(x, x') \mapsto H(x, z, x', v)$ is
  strictly concave and the second inequality holds because
  $\theta\sigma^*(x, z) + (1-\theta) \sigma^*(y, z) \in \Gamma(x_\theta,
  z)$ by Assumption~\ref{a:ibc}. The strict
  concavity of $H$ and the maximum theorem imply that $x \mapsto
  \sigma^*(x, z)$ is single-valued and continuous.

  Now we add Assumption~\ref{a:diff} and consider differentiability.
  Since $\sigma^*(x_0, z_0)\in \interior\Gamma(x_0, z_0)$ and $\Gamma$ is 
  continuous, there exists an open neighborhood $O$ of $x_0$
  such that $\sigma^*(x_0, z_0) \in \interior\Gamma(x, z_0)$ for all $x\in
  O$. On $O$ we define $W(x) := H\left(x, z_0, \sigma^*(x_0, z_0),
    v^*\right)$. Then $W(x) \leq v^*(x, z_0)$ on $O$ and $W(x_0) = v^*(x_0,
  z_0)$. The claim follows then from 
  Assumption~\ref{a:diff} and \cite{benveniste1979differentiability}.
\end{proof}

\subsection{Proofs for Section~\ref{s:lrc}}\label{s:lrc-proofs}

\begin{proof}[Proof of Proposition~\ref{p:0}]
  Since $\beta \in bc\ZZ$, $L_\beta$ defined in \eqref{eq:B} is a
  bounded linear operator. It follows from Theorem~1.5.5 of
  \cite{buhlerfunctional} that $r(L_\beta) :=
  \lim_{n\to\infty}\|L_\beta^n\|^{1/n}$ always exists and is bounded above
  by $\|L_\beta\| = \sup_z\beta(z)$.
  
  Let $\mathbbm{1} \equiv 1$ on $\ZZ$. For each $z\in \ZZ$ and $n\in \NN$, an inductive argument gives
  \begin{equation}
    \label{eq:r}
    \EE_z \prod_{t=0}^{n-1}\beta(Z_t) = L_\beta^n \mathbbm{1}(z).
  \end{equation}
  Thus, eventual discounting can be written
  as $\|L_\beta^n\mathbbm{1}\| < 1$ for some $n\in \NN$.
    Applying Theorem~9.1 of \citet{krasnoselskii1972approximate},  since
    (i) $L_\beta$ is a positive linear operator on $bm\ZZ$, (ii) the positive
    cone in this set is solid and normal under the pointwise partial order\footnote{A cone is
    solid if it has an interior point; it is normal if $0 \leq x \leq y$
    implies that $\|x\| \leq M \|y\|$. The cone of nonnegative functions in
    $bm\ZZ$ is both solid and normal.}, and (iii) $\1$ lies interior to the
    positive cone in $bm\ZZ$, we have
    \begin{equation}
        \label{eq:arlb}
        r(L_\beta) 
         = \lim_{n\to\infty} \|L_\beta^n \1 \|^{1/n}
         = \lim_{n\to\infty} \left\{ \sup_{z \in \ZZ} \EE_z \prod_{t=0}^{n-1}\beta(Z_t) \right\}^{1/n} ,
    \end{equation}
  where the second equality is due to \eqref{eq:r}, nonnegativity of $\beta$
  and the definition of the supremum norm.   This confirms the first claim in
  Proposition~\ref{p:0}.  It also follows immediately that
  $r(L_\beta) < 1$ implies eventual discounting.

  To see that the converse is true, suppose there exists an $n\in\NN$ such that
  $r_n^\beta < 1$. Then any $m\in\NN$ can be expressed uniquely as $m =
  kn + i$ for some $k, i\in\NN$ with $i<n$. For sufficiently large $m$, it
  follows from the Markov property that
  \begin{align*}
      \left(r_m^\beta\right)^{1/m}
      &= \left\{ \sup_{z \in \ZZ} \EE_z \prod_{t=0}^{n-1}\beta(Z_t)
          \left[
              \EE_{Z_{n-1}} \prod_{t=n}^{m-1} \beta(Z_t)
          \right]
      \right\}^{1/m}\\
      &\leq \big( r_n^\beta\, r_{m-n}^\beta \big)^{1/m}
      \leq \big( r_n^\beta \big)^{k/m} \big( r_i^\beta \big)^{1/m}.
  \end{align*}
  The right hand side is dominated by $(r_n^\beta)^{k/m} M^{1/m}$, where $M
  := \sup_{i<n}r_i^\beta < \infty$.
  If $m \to \infty$, then $k/m \to 1/n$, and this term approaches
  $(r_n^\beta)^{1/n} < 1$. Hence $r(L_\beta) < 1$, as was to be shown.
\end{proof}

\begin{proof}[Proof of Proposition~\ref{p:fic}]
    The proof of Proposition~\ref{p:0} uses the fact that
    $r_\BB(M) = \lim_{n\to\infty} \|M^n h \|_\BB^{1/n}$ holds
    when $M$ is a positive (i.e., order preserving) linear operator on a Banach lattice $(\BB, \| \cdot
    \|_\BB)$ with solid positive cone, $r_\BB$ denotes the spectral radius
    of a linear operator mapping this Banach lattice to itself, and
    $h$ is interior to the positive cone
    \citep[Theorem~9.1]{krasnoselskii1972approximate}.
    If $\ZZ$ is finite and $\{Z_t\}$ is irreducible with stationary
    distribution $\pi$, we can take $\BB$ to be all $h \colon \ZZ \to \RR$ and
    set $\| h \|_\BB = \sum_{z \in \ZZ} |h(z)| \pi(z) =: \EE_\pi
    h$. Under this norm, $\1$ is interior to the positive cone of $\BB$
    because, by irreducibility, $\pi(z) > 0$ for all $z \in \ZZ$.
    Applying the above expression for the spectral radius to $L_\beta$, as
    well as the result in \eqref{eq:r}, we obtain
    \begin{equation}
        r_\BB(L_\beta) 
        = \lim_{n\to\infty} \|L_\beta^n \1 \|_\BB^{1/n}
        = \lim_{n\to\infty}
            \left\{ 
              \EE_\pi  \EE_z \prod_{t=0}^{n-1}\beta(Z_t)
            \right\}^{1/n}    
            = \lim_{n\to\infty} (s^\beta_n)^{1/n},
    \end{equation}
    where the last equality uses the law of iterated expectations and the
    definition of $s^\beta_n$ in Proposition~\ref{p:fic}.  

    It remains only to show that $r_\BB(L_\beta) = r(L_\beta)$, where
    the latter is defined, as before, using the supremum norm (see, e.g.,
    \eqref{eq:arlb}).  In other words, we need to show that 
    \begin{equation}\label{eq:eqlim}
        \lim_{n\to\infty} \|L_\beta^n \1 \|^{1/n} = \lim_{n\to\infty} \|L_\beta^n \1 \|_\BB^{1/n}.    
    \end{equation}
    On finite dimensional normed linear spaces, any two norms are
    equivalent (see, e.g., \cite{buhlerfunctional}, Theorem~1.2.5),
    so we can take positive constants $c$ and $d$ with $\| \cdot \| \leq
    c \| \cdot \|_\BB \leq d \| \cdot \|$ on $\BB$.
    The equality in \eqref{eq:eqlim} easily follows and 
    the proof is now complete.
\end{proof}

\subsection{Proofs for Section~\ref{s:ubr}}

\subsubsection{Homogeneous Functions}

Let the operators $T_\sigma$ and $T$ be as defined in~\eqref{eq:T_sigma} and
\eqref{eq:bellman}, respectively, with aggregator $H$ given by~\eqref{eq:H}.
The definition of $h_\theta \sS$ is given in Section~\ref{ss:hom}.

\begin{proof}[Proof of Proposition~\ref{p:hom}]
  We first show that $T$ is eventually contracting on $\vV=h_\theta \sS$.
  Since Assumption~\ref{a:hom1} holds, the Feller property implies that $T$ maps $\vV$ to itself. Note that
  for any $v\in \vV$, we have $v(x, z) = |x|^\theta v(x/|x|, z)$. It
  follows from Assumption~\ref{a:hom2} that for any $v, w\in \vV$,
  \begin{align*}
    &\phantom{\leq\;} \left|(T^n v)(x_0, z_0) - (T^n w)(x_0, z_0)\right|\\
    &\leq \sup_{x_1\in \Gamma(x_0, z_0)}\beta(z_0) \int
    \left|(T^{n-1} v)(x_1, z_1) - (T^{n-1} w)(x_1, z_1)\right|Q(z_0, dz_1)\\
    &\leq \sup_{x_1\in\Gamma(x_0, z_0)}\beta(z_0) \int |x_1|^\theta
    \left|(T^{n-1} v)\left(\frac{x_1}{|x_1|}, z_1\right) -
      (T^{n-1} w)\left(\frac{x_1}{|x_1|}, z_1\right)\right|Q(z_0, dz_1)\\
    &\leq \sup_{x_1\in\Gamma(x_0, z_0)} \beta(z_0) \alpha^\theta(z_0)
    |x_0|^\theta \int \left|(T^{n-1} v)\left(\frac{x_1}{|x_1|},
        z_1\right) - (T^{n-1} w)\left(\frac{x_1}{|x_1|},
        z_1\right)\right|Q(z_0, dz_1).
  \end{align*}
  An inductive argument gives that
  \begin{align*}
    &\phantom{\leq\;} \left|(T^n v)(x_0, z_0) - (T^n w)(x_0, z_0)\right|\\
    &\leq |x_0|^\theta \sup_{x_1\in \Gamma(x_0, z_0)} \EE_{z_0}
    \prod_{t=0}^{n-1} \beta(z_t)\alpha^\theta(z_t)
    \left|v\left(\frac{x_n}{|x_n|}, z_n\right) -
      w\left(\frac{x_n}{|x_n|}, z_n\right)\right|\\
    &\leq |x_0|^\theta \left(\EE_{z_0} \prod_{t=0}^{n-1}
      \beta(z_t)\alpha^\theta(z_t) \right) \|v - w\|_h
  \end{align*}
  where the norm $\|\cdot\|_h$ is defined in \eqref{eq:Hnorm}. Therefore, we
  have
  \begin{equation*}
    \|T^n v - T^n w\|_h \leq \sup_{z_0\in\ZZ}\left(\EE_{z_0}
      \prod_{t=0}^{n-1} \beta(z_t)\alpha^\theta(z_t) \right) \|v-w\|_h.
  \end{equation*}
  By Assumption~\ref{a:hom2}, $T$ is eventually contracting on $\vV$. Hence,
  $T$ has a unique fixed point $\bar{v}$ on $\vV$ and $T^n v \to \bar{v}$
  for any $v\in \vV$.

  Since $T_\sigma v$ is not necessarily in $\vV$, we cannot apply the same
  argument to $T_\sigma$. Hence, we prove the remaining results directly. We
  first show that $v_\sigma := \lim_n (T^n_\sigma \textbf{0})$ is well
  defined. It follows from Assumptions~\ref{a:hom1} and \ref{a:hom2} that
  \begin{align*}
    (T^n_\sigma \textbf{0})(x_0, z_0) &= \sum_{t=0}^{n-1} \EE_{z_0}
    \prod_{i=0}^{t-1} \beta(z_i) u(x_t, z_t, \sigma(x_t, z_t)) \\
    &\leq \sum_{t=0}^{n-1} \EE_{z_0} \prod_{i=0}^{t-1} \beta(z_i) \left|u(x_t,
      z_t, \sigma(x_t, z_t))\right|
    \\
    &\leq \sum_{t=0}^{n-1} \EE_{z_0} \prod_{i=0}^{t-1} \beta(z_i)
    \alpha(z_i)^\theta
    B(1+\alpha(z_t))^\theta|x_0|\\
    &\leq \sum_{t=0}^{n-1} \EE_{z_0} \prod_{i=0}^{t-1} \beta(z_i)
    \alpha(z_i)^\theta B(1+\bar{\alpha})^\theta|x_0|
  \end{align*}
  where $\bar{\alpha} = \sup_{z\in\ZZ}\alpha(z)$. It follows from
  Proposition~\ref{p:0} and the Cauchy root test that the series converges
  absolutely and hence $v_\sigma(x_0, z_0)$ is finite and well defined.

  Next we show that $\bar{v} = v^*$. Since $\bar{v} = T\bar{v}$, we have for
  any $\sigma\in \Sigma$,
  \begin{align*}
      \bar{v}(x_0, z_0) &= \max_{x'\in \Gamma(x_0, z_0)}
      \left\{
          u(x_0, z_0, x') + \beta(z_0)\int \bar{v}(x', z_1)Q(z_0, dz_1)
      \right\} \\
      &\geq u(x_0, z_0, \sigma(x_0, z_0)) + \beta(z_0) \int
      \bar{v}(\sigma(x_0, z_0), z_1)Q(z_0, dz_1).
  \end{align*}
  It follows from induction that
  \begin{equation}
    \label{eq:v_upper}
    \bar{v}(x_0, z_0) \geq (T^n_\sigma \textbf{0})(x_0, z_0) + \EE_{z_0}
    \prod_{t=0}^{n-1} \beta(z_t) \bar{v}(x_n, z_n)
  \end{equation}
  where $\{x_n\}$ is given by $\sigma$. Since $\bar{v}\in\vV$, we have
    $\bar{v}(x_n, z_n) \leq \prod_{t=0}^{n-1} \alpha(z_t)^\theta
    |x_0|^\theta \|\bar{v}\|_h$.
  Taking $n$ to infinity in \eqref{eq:v_upper}, the last term goes to 0 and thus
  $\bar{v} \geq v_\sigma$ for all $\sigma\in\Sigma$. By the
  measurable maximum theorem, we can find $\sigma^*\in\Sigma$ such that
  $T\bar{v} = T_{\sigma^*} \bar{v}$. A similar argument shows that
  $\sigma^*$ achieves the maximum. Therefore, $\bar{v}$ is the value
  function and $\sigma^*$ is the optimal policy.

  Because $v^* = T_{\sigma^*}v^*$ is homogeneous of degree $\theta$, we have
  for any $\lambda \geq 0$,
  \begin{equation*}
    v^*(\lambda x, z) 
    = \lambda^\theta v^*(x, z)
    = \lambda^\theta u(x, z, \sigma^*(x, z)) + \beta(z)\int \lambda^\theta
    v^*(\sigma^*(x, z), z')Q(z, dz').
  \end{equation*}
  It follows that $\sigma^*(\lambda x, z) = \lambda \sigma^*(x, z)$, that is,
  the optimal policy is homogeneous of degree one.
\end{proof}

\subsubsection{Local Contractions}

Recall that the operators $T_\sigma$ and $T$ are as defined
in~\eqref{eq:T_sigma} and \eqref{eq:bellman}, respectively, with aggregator
$H$ given by~\eqref{eq:H}.

\begin{proof}[Proof of Proposition~\ref{p:loc}]
  Define
    $u_j(x, z) := \max_{x'\in \Gamma(x, z)}|u(x, z, x')|$  if $x\in
    K_j$ and $r_j := \sup_{x\in K_j, z\in Z} u_j(x, z)$.
  Since $u$ is continuous and every $K_j$ is compact, $r_j<\infty$ for all
  $j$. For any initial state $(x_0, z_0)$, we can find $j$ such that $x_0\in
  K_j$. It follows from Assumption~\ref{a:loc2} that
    $|u(x_t, z_t, x_{t+1})| \leq
    r_j$ for all $t\in \NN$.

  Choose any increasing and unbounded $\{m_j\}$ such that $m_j \geq r_j$.
  Since $Q$ is Feller, $Tv$ is continuous on every $K_j$ for $v\in c_m\sS$,
  where the space $c_m\sS$ is defined in Section~\ref{ss:loc}. It follows
  from Remark 1(a) of \citet{matkowski2011discounted} that $T: c_m\sS \to
  c\sS$.
  
  Since $\Gamma(x, z) \subset K_j$ for all $x\in K_j$, we have on $K_j$
  \begin{align*}
    |(T^nv)(x, z) - (T^nw)(x, z)| &\leq \sup_{x'\in \Gamma(x, z)} \beta(z)
    \int |T^{n-1}v(x', z') - T^{n-1}w(x', z')| Q(z, dz')\\
    &\leq \sup_{x'\in K_j} \beta(z) \int |T^{n-1}v(x', z') -
    T^{n-1}w(x', z')| Q(z, dz')\\
    &\leq \beta(z) \|T^{n-1} v - T^{n-1}w\|_j.
  \end{align*}
  An inductive argument gives
  \begin{equation*}
    |(T^nv)(x, z) - (T^nw)(x, z)| \leq \EE_z \prod_{t=0}^{n-1}\beta(Z_t) \|v
    - w\|_j.
  \end{equation*}
  Taking the supremum, we have $\|T^nv - T^nw\|_j \leq r_n^\beta \|v-w\|_j$.
  Since $(\beta, Q)$ is eventually discounting, $T^n$ is a 0-local
  contraction for some $n\in\NN$.\footnote{We say an 
    operator $T:c_m\sS \to c\sS$ is a 0-local contraction if there
    exists a $\beta\in (0, 1)$ such that $\|Tf - Tg\|_j \leq \beta \|f -
    g\|_j$ for all $f, g\in c_m\sS$ and all $j \in \NN$.} Then it follows
  from Proposition 1 of 
  \citet{matkowski2011discounted} that $T$ has a unique fixed point
  $\bar{v}$ in $c_m\sS$. It can be proved in the same way that $T_\sigma^n$
  is also a 0-local contraction and hence $v_\sigma$ is well defined and
  finite for any initial state. Since we can find $\sigma$ such that
  $T_\sigma \bar{v} = T\bar{v}$ by the measurable maximum theorem, the
  optimality results follow from a similar argument to the proofs of
  Theorem~\ref{t:bk}.
\end{proof}

\subsection{Proofs for Section~\ref{s:ext}}

\subsubsection{Alternative Discount Specifications}

Here we sketch the proof of Theorem~\ref{t:bk} for the alternative timing
when the aggregator satisfies \eqref{eq:bc2}. Let $\{Z_t\}$ be a
Markov process generated by $Q$ starting at $z = Z_0$
and let $\beta_t = \beta(Z_{t+1})$. A similar argument to the proof
of Lemma~\ref{l:T_sigma} yields
    $|T^n_\sigma v - T^n_\sigma w|
    \leq \EE_{z} \prod_{t=1}^n
    \beta(Z_{t+1}) \|v - w\|$,
where $\EE_{z}$ represents expectation conditional on $Z_0 = z$. Taking the
supremum gives $\|T^n_\sigma v - T^n_\sigma w\| \leq 
r_n^\beta \|v - w\|$. Similar result holds for the
Bellman operator $T$. Therefore, both $T_\sigma$ and $T$ are eventually
contracting if $r_n^\beta < 1$ for some $n \in \NN$.
The rest of the proof remains the same.

\begin{proof}[Proof of Proposition~\ref{p:shocks}]
    Recall that the primitives are redefined as in footnote~\ref{fn:redef}.
    Then the aggregator satisfies
    \begin{equation*}
        |H(x, z, x', v) - H(x, z, x', w)| 
        \leq \int \beta(z') |v(x', z') - w(x', z')|\tilde{Q}(z, dz').
    \end{equation*}
    Based on the discussion above, the eventual discounting condition
    remains the same. It then follows from Proposition~\ref{p:0} that
    eventual discounting holds if and only if $r(L_\beta) < 1$ and
    \begin{equation*}
        r(L_\beta) = \lim_{n\to\infty} (r^\beta_n)^{1/n} = \lim_{n\to\infty}
        \left(
            \sup_{z \in \tilde{\ZZ}} \tilde{\EE}_{z} \prod_{t=1}^n
            \beta(\tilde{Z}_{t+1})
        \right)^{1/n}
    \end{equation*}
    where $\tilde{\EE}_z$ represents conditional expectation under
    $\tilde{Q}$. Since $\tilde{Q}$ is induced by $Q$ and
    $\beta(\tilde{Z}_{t+1}) = bZ_{t+1}/Z_t$, we can write $r^\beta_n =
    \sup_{z\in\ZZ}\EE_z b^n Z_t$. Then we have
    $(b^n z_a)^{1/n} \leq (r^\beta_n)^{1/n} \leq(b^n z_b)^{1/n}$,
    where $z_a$ and $z_b$ are 
    positive constants such that $z_a < Z_t < z_b$ for all $t$. 
    Taking $n \to \infty$ gives $r(L_\beta) = b$, so eventual
    discounting holds if and only if $b<1$.
\end{proof}


\subsubsection{Epstein-Zin Preferences}
\label{ss:EZ-proofs}

For ease of notation, we replace $1/\psi$ with $\rho$ in what follows.
The definition of $\vV$ and 
$\|f\|_I$ are given in Section~\ref{ss:ez}.
Let the operators $T$ and $T_\sigma$ be as defined in \eqref{eq:T_sigma} and
\eqref{eq:bellman}, respectively, with aggregator $H$ given by
\eqref{eq:EZ-H}. Let $\tilde{T}_\sigma$ and $\tilde{T}$ be defined in the
same way except that $H$ is replaced by
\begin{equation}
  \label{eq:EZ-H2}
  \tilde{H}(x, z, c, v) = \left\{c^{1-\rho} + \beta(z) \left[ \int
      v\left(R(z)(x-c), z'\right) Q(z, dz') \right]^{1-\rho}
  \right\}^{\frac{1}{1-\rho}},
\end{equation}
which is a special case of $H$ when $\gamma=0$. We first prove a useful
lemma.

\begin{lemma}
  \label{l:jensen}
  $T_\sigma v \leq \tilde{T}_\sigma v$ and $T v \leq \tilde{T} v$ for all
  $v\in\vV$.
\end{lemma}
\begin{proof}
  Since $\gamma > 1$, by Jensen's inequality, we have 
  \begin{equation*}
    \left[ \int v^{1-\gamma}(x, z') Q(z, dz')
    \right]^{\frac{1}{1-\gamma}} \leq \int v(x, z') Q(z, dz')
  \end{equation*}
  for all $(x, z) \in \sS$ and $v\in\vV$. It follows that
  \begin{align*}
    (T_\sigma v)(x, z) &\leq \left\{\sigma(x, z)^{1-\rho} + \beta(z) \left[
        \int v\left[R(z)\left(x-\sigma(x, z)\right), z'\right] Q(z, dz')
      \right]^{1-\rho} \right\}^{\frac{1}{1-\rho}}\\
    &= (\tilde{T}_\sigma v)(x, z).
  \end{align*}
  That $T v \leq \tilde{T} v$ can be shown in a similar way.
\end{proof}

A central result of this section is the following proposition, which
guarantees that the $\sigma$-value function $v_\sigma =
\lim_n(T_\sigma^n\textbf{0})$ is well defined and a fixed point of
$T_\sigma$.

\begin{proposition}
  \label{p:v_sigma}
  Under Assumption~\ref{a:EZ}, there exists a function $\hat{v}: \sS \to
  \RR_+$ given by
  \begin{equation}
    \label{eq:vhat}
    \hat{v}(x_0, z_0) := x_0\left\{\lim_{n\to\infty} \sum_{t=0}^{n-1}
      \left[\EE_{z_0} \prod_{i=0}^{t-1} \beta(z_i)^{\frac{1}{1-\rho}} R(z_i)
      \right]^{1-\rho}\right\}^{\frac{1}{1-\rho}}
  \end{equation}
  %
  such that $\hat{v}\in\vV$ and $T_\sigma$ is a self map on $[0, \hat{v}]
  \subset \vV$. The $\sigma$-value function is well defined and is the least
  fixed point of $T_\sigma$ on $[0, \hat{v}] \subset \vV$. Furthermore, if
  $\sigma$ satisfies that $\inf_{z\in\ZZ}\sigma(x, z)/x > 0$ for all $x>0$,
  then $v_\sigma$ is the unique fixed point of $T_\sigma$ on $[0, \hat{v}]
  \subset \vV$ and $T^n_\sigma v \to v_\sigma$ for all $v\in [0, \hat{v}]
  \subset \vV$.
\end{proposition}

We first give two lemmas that are crucial to the proof of
Proposition~\ref{p:v_sigma}. The first lemma shows that $\hat{v}$ can indeed
act as an upper bound function.

\begin{lemma}
  \label{l:ubound}
  $\hat{v} \in \vV$ and $T_\sigma \hat{v} \leq \hat{v}$ for all $\sigma \in
  \Sigma$.
\end{lemma}

\begin{proof}
  Let $\hat{v}_n(x_0, z_0) := x_0 A_n(z_0)^{1/(1-\rho)}$ where
  \begin{equation*}
    A_n(z_0) := \sum_{t=0}^{n-1} \left[\EE_{z_0} \prod_{i=0}^{t-1}
      \beta(z_i)^{\frac{1}{1-\rho}} R(z_i) \right]^{1-\rho}.
  \end{equation*}
  By Proposition~\ref{p:0} and Assumption~\ref{a:EZ}, we have
  \begin{equation*}
    \limsup_{n\to\infty} \left[\sup_{z_0\in\ZZ}\EE_{z_0} \prod_{i=0}^{t-1}
      \beta(z_i)^{\frac{1}{1-\rho}} R(z_i) \right]^{\frac{1-\rho}{n}} =
    r(L_R)^{1-\rho}< 1,
  \end{equation*}
  where $L_R$ is as defined in \eqref{eq:lrdef}.
  It follows from the root test that $\lim_nA_n$ is well defined and bounded
  on $\ZZ$. Hence, $\hat{v} = \lim_n \hat{v}_n$ and it satisfies
    $\|\hat{v}\|_I = \sup_{x\in\XX, z\in\ZZ} |x A(z)/(1+x) | \leq
    \sup_{z\in\ZZ} A(z) < \infty$.
  Therefore, $\hat{v} \in \vV$.

  Next, we use the operator $\tilde{T}_\sigma$ defined above to show that
  $T_\sigma \hat{v} \leq \hat{v}$. Since $A_n$ is increasing in $n$, by the
  Monotone Convergence Theorem, we have
    $\lim_{n\to\infty} (\tilde{T}_\sigma \hat{v}_n)(x_0, z_0) =
    (\tilde{T}_\sigma \hat{v})(x_0, z_0)$.
  Write $A_n(z_0) = \sum_{t=0}^{n-1} B_t(z_0)$. Since $\sigma(x, z) \leq x$, it
  follows that
  \begin{align*}
    (\tilde{T}_\sigma \hat{v}_n)(x_0, z_0) &\leq x_0\left\{1 +
      \left[\beta(z_0)^{\frac{1}{1-\rho}}R(z_0) \EE_{z_0}
        A_n(z_1)^{\frac{1}{1-\rho}}
      \right]^{1-\rho} \right\}^{\frac{1}{1-\rho}}\\
    &= x_0\left\{1 + \left[\beta(z_0)^{\frac{1}{1-\rho}}R(z_0) \EE_{z_0}
        \left( \sum_{t=0}^{n-1} B_t(z_1) \right)^{\frac{1}{1-\rho}}
      \right]^{1-\rho} \right\}^{\frac{1}{1-\rho}}.
  \end{align*}
  Since $\rho \in (0, 1)$, by the Minkowski inequality, we have
  \begin{equation*}
    (\tilde{T}_\sigma \hat{v}_n)(x_0, z_0) \leq x_0\left\{1 +
      \sum_{t=0}^{n-1} \left[\beta(z_0)^{\frac{1}{1-\rho}}R(z_0) \EE_{z_0}
        B_t(z_1)^{\frac{1}{1-\rho}} \right]^{1-\rho}
    \right\}^{\frac{1}{1-\rho}}.
  \end{equation*}
  Note that the following equation holds
  \begin{equation*}
    \beta(z_0)^{\frac{1}{1-\rho}}R(z_0)\EE_{z_0}B_t(z_1)^{\frac{1}{1-\rho}}
    = B_{t+1}(z_0)^{\frac{1}{1-\rho}}
  \end{equation*}
  by the Markov property. It follows that
  \begin{equation*}
    (\tilde{T}_\sigma \hat{v}_n)(x_0, z_0) 
    \leq x_0\left\{ 1 + \sum_{t=1}^n B_t(z_0) \right\}^{\frac{1}{1-\rho}}
    = x_0 A_{n+1}(z_0)^{\frac{1}{1-\rho}} 
    = \hat{v}_{n+1}(x_0, z_0).
  \end{equation*}
  Taking $n$ to infinity, we have $\tilde{T}_\sigma \hat{v} \leq \hat{v}$.
  By Lemma~\ref{l:jensen}, $T_\sigma \hat{v} \leq \hat{v}$.
\end{proof}

\begin{lemma}
  \label{l:self}
  $T_\sigma v \in \vV$ for all $\sigma\in\Sigma$ and $v\in\vV$.
\end{lemma}

\begin{proof}
  Evidently $T_\sigma v$ is measurable given $\sigma\in \Sigma$. To
  see that $T_\sigma v$ is bounded, we have
  \begin{align*}
    (T_\sigma v)(x, z) &\leq \left\{\sigma(x, z)^{1-\rho} + \beta(z) \left[
        \int v\left[R(z)\left(x-\sigma(x, z)\right), z'\right] Q(z, dz')
      \right]^{1-\rho} \right\}^{\frac{1}{1-\rho}}\\
    &\leq \left\{x^{1-\rho} + \beta(z) \|v\|_I^{1-\rho}\left[1 +
        R(z)x\right]^{1-\rho}
    \right\}^{\frac{1}{1-\rho}},
  \end{align*}
  where the first inequality follows from Lemma~\ref{l:jensen} and the
  second inequality follows from the fact that $\sigma(x, z) \in [0, x]$ and
  $|v(x, z)| \leq \|v\|_I (1+x)$ for all $v\in\vV$. Dividing both sides by
  $(1+x)$ yields (assuming $\sup_{z}R(z) > 1$)
  \begin{equation*}
    \|T_\sigma v\|_I \leq \sup_{z\in\ZZ} \left\{ 1 + \beta(z)
      \|v\|_I^{1-\rho} R(z)^{1-\rho} \right\}^{\frac{1}{1-\rho}}.
  \end{equation*}
  Since $\beta$ and $R$ are
  bounded, $\|T_\sigma v\|_I < \infty$.
\end{proof}

\begin{proof}[Proof of Proposition~\ref{p:v_sigma}]
  It is apparent that $T_\sigma \textbf{0} \geq \textbf{0}$. It follows from
  Lemma~\ref{l:ubound}, Lemma~\ref{l:self}, and the monotonicity of
  $T_\sigma$ that $T_\sigma$ is a self map on $[0, \hat{v}] \subset \vV$.
  Let $\{v_n\}$ be a countable chain\footnote{A set $C \subset \vV$ is
    called a chain if for every $x, y\in C$, either $x\leq y$ or $y \leq
    x$.} on $[0, \hat{v}] \subset \vV$. Then both $\sup_n v_n$ and $\inf_n
  v_n$ are measurable and bounded in norm by $\|\hat{v}\|_I$. So $[0,
  \hat{v}] \subset \vV$ is a countably chain complete partially ordered set.
  For any increasing $\{v_n\} \subset [0, \hat{v}]$, it follows from the
  Monotone Convergence Theorem that $\sup_n T_\sigma v_n = T_\sigma (\sup_n
  v_n)$. Hence, $T_\sigma$ is monotonically sup-preserving. Then, by the
  Tarski-Kantrovich Theorem,\footnote{See, for example,
    \cite{becker2018recursive} for a version of the theorem and related
    definitions.} $v_\sigma := \lim_n (T_\sigma \textbf{0})$ is the least
  fixed point of $T_\sigma$ on $[0, \hat{v}] \subset \vV$.

  If $\sigma$ satisfies that $\inf_{z\in\ZZ}\left(\sigma(x, z)/x\right) > 0$
  for all $x>0$, then there exists an $\alpha > 0$ such that $\sigma(x, z) \geq
  \alpha x \sup_zA(z) \geq \alpha \hat{v}(x, z)$. Since $T_\sigma \textbf{0}
  = \sigma \leq \hat{v}$, $T_\sigma \textbf{0}$ and $\hat{v}$ are
  comparable. Uniqueness and convergence then follow from Theorems~10 and 11
  in \cite{marinacci2010unique}.
\end{proof}

Recall from Section~\ref{ss:ez} that $\hat{\vV}$ is all functions in $\vV$
that are homogeneous of degree one in $x$. The following lemma is useful in
the proof of Proposition~\ref{p:vstar}.

\begin{lemma}
  \label{l:self_T}
  For any $v\in\hat{\vV}$, $T v \in \hat{\vV}$ and there exists a
  $\sigma\in\Sigma$ homogeneous in $x$ that satisfies $Tv = T_\sigma v$ and
  $\inf_z\sigma(x, z)/x > 0$ for all $x>0$.
\end{lemma}

\begin{proof}
  Pick $v\in\hat{\vV}$ and we can write $v(x, z) = x h(z)$ for some bounded
  measurable $h$. Then \eqref{eq:EZ-H} becomes
  \begin{equation}
    \label{eq:EZ-H3}
    H(x, z, c, v) = \left\{c^{1-\rho} + \beta(z)R(z)^{1-\rho}(x-c)^{1-\rho}
      \left[ \int h(z')^{1-\gamma} Q(z, dz')
      \right]^{\frac{1-\rho}{1-\gamma}} \right\}^{\frac{1}{1-\rho}}.
  \end{equation}
  Since $c \mapsto H(x, z, c, v)$ is continuous and $(x, z) \mapsto H(x, z,
  c, v)$ is measurable, by the measurable maximum theorem, $Tv$ is
  measurable and there exists a $\sigma\in\Sigma$ such that $T_\sigma v = T
  v$. Since $c \leq x$ in~\eqref{eq:EZ-H3}, a similar argument to the proof
  of Lemma~\ref{l:self} shows that $Tv$ is bounded in $\|\cdot\|_I$.

  In fact, $\sigma(x, z)$ is the solution of the single variable
  optimization problem maximizing $ c^{1-\rho} + (x-c)^{1-\rho}f(z)$ over 
    $0\leq c\leq x$ where
  \begin{equation*}
    f(z) := \beta(z)R(z)^{1-\rho} 
        \left[ \int h(z')^{1-\gamma} Q(z, dz') \right]^{\frac{1-\rho}{1-\gamma}}.
  \end{equation*}
  It has closed-form solution 
    $\sigma(x, z) = x/(f(z)^{1/\rho} + 1)$.
  Therefore, $\sigma$ is homogeneous in $x$ and thus $Tv = T_\sigma v$ is
  also homogeneous in $x$. It follows that $Tv \in \hat{\vV}$. Since $f(z)$
  is bounded, $\inf_z\sigma(x, z)/x > 0$.
\end{proof}

\begin{proof}[Proof of Proposition~\ref{p:vstar}]
  By Lemma~\ref{l:self_T}, there exists a $\sigma$ such that $T_\sigma \hat{v}
  = T\hat{v}$. It follows from Lemma~\ref{l:ubound} that $T_\sigma \hat{v}
  \leq \hat{v}$ and hence $T\hat{v} \leq \hat{v}$. Then the monotonicity of
  $T$ implies that $Tv \leq \hat{v}$ for all $v\in\hat{\vV}$. By
  Lemma~\ref{l:self_T} and the monotonicity of $T$, $T^n \textbf{0}$ is an
  increasing sequence on $\hat{\vV}$ bounded above by $\hat{v}$. Therefore,
  the pointwise limit $\bar{v} := \lim_{n\to\infty} (T^n\textbf{0})$ is well
  defined and is also in $[0, \hat{v}] \subset \hat{\vV}$.
  

  To see that $\bar{v}$ is the value function, pick any $\sigma\in\Sigma$.
  Since $T^n\textbf{0}$ is an increasing sequence converging to $\bar{v}$,
  $\bar{v} \geq T^n\textbf{0} \geq T^n_\sigma \textbf{0}$. Taking $n$ to
  infinity, it follows from Proposition~\ref{p:v_sigma} that $\bar{v} \geq
  v_\sigma$. Next we show that $\bar{v}$ can be achieved by a feasible
  policy. Since $T^n \textbf{0} \leq \bar{v}$, the monotonicity of $T$
  implies that $T^{n+1}\textbf{0} \leq T\bar{v}$. Taking $n$ to infinity
  yields $\bar{v} \leq T\bar{v}$. By Lemma~\ref{l:self_T}, there exists a
  homogeneous $\sigma^*\in\Sigma$ that satisfies the interiority condition
  and $T_{\sigma^*} \bar{v} = T\bar{v}$. Then we have $\bar{v} \leq
  T_{\sigma^*} \bar{v}$ and hence $\bar{v} \leq T_{\sigma^*}^n \bar{v}$ by
  the monotonicity of $T_{\sigma^*}$. Taking $n$ to infinity, it follows
  from Proposition~\ref{p:v_sigma} that $\bar{v} \leq v_{\sigma^*}$. Since
  $\bar{v} \geq v_\sigma$ for all $\sigma \in \Sigma$, $\bar{v} =
  v_{\sigma^*}$.
\end{proof}

For the specification in \cite{de2018uncertainty} where the lifetime utility
satisfies
\begin{equation*}
    U(C_t, C_{t+1}, \ldots) = \left\{ (1-\beta_t) C_t^{1-\rho} + \beta_t
      \left[
        \EE_t U^{1-\gamma}(C_{t+1}, C_{t+2}, \ldots)
      \right]^{\frac{1-\rho}{1-\gamma}} \right\}^{\frac{1}{1-\rho}},
\end{equation*}
we can redefine the upper bound function to be
\begin{equation*}
    \tilde{v}(x_0, z_0) := x_0
    \left\{
      \lim_{n\to\infty} \sum_{t=0}^{n-1}
      \left[
        \EE_{z_0} \prod_{i=0}^{t-1} \beta(z_i)^{\frac{1}{1-\rho}} R(z_i)
        \left[1 - \beta(z_t) \right]^{\frac{1}{1-\rho}}
      \right]^{1-\rho}
    \right\}^{\frac{1}{1-\rho}}.
\end{equation*}
Since $\beta(z_t) < 1$, $\tilde{v}$ is bounded above by $\hat{v}$ in
\eqref{eq:vhat}. Then it can be shown that all the above results hold for
the new preference if Assumption~\ref{a:EZ} is satisfied. The proof is
omitted.

\subsection{Analytical Expression for the Geometric Mean}
\label{s:geomean}

Consider $\beta_t = \exp(\alpha Z_t)$ where $\{Z_t\}$ obeys \eqref{eq:ar1}.
An inductive argument shows that for all $t\geq 1$,
\begin{equation}\label{eq:zt}
    Z_t = (1-\rho^t)\mu + \rho^t Z_0 + \sigma_\epsilon
    (\epsilon_t + \rho \epsilon_{t-1} + \ldots + \rho^{t-1}\epsilon_1).
\end{equation}
It follows that
\begin{equation*}
    \sum_{t=0}^{n-1} Z_t
    = \left( n - \frac{\rho(1-\rho^n)}{1-\rho} \right)\mu +
    \frac{1-\rho^{n+1}}{1-\rho} Z_0 +
    \sigma_\epsilon\left( \epsilon_n + \frac{1-\rho^2}{1-\rho}\epsilon_{n-1}
    + \ldots + \frac{1-\rho^n}{1-\rho}\epsilon_1 \right).
\end{equation*}
Exploiting the properties of log-normal distributions, we have
\begin{equation*}
    \EE_z \exp\left(\sum_{t=0}^{n-1} Z_t \right)
    = \exp\left(
      n\mu - \frac{\rho(1-\rho^n)}{1-\rho}\mu +
      \frac{1-\rho^{n+1}}{1-\rho}z +
      \frac{\sigma_\epsilon^2}{2}\sum_{t=1}^n m_t
    \right)
\end{equation*}
where $m_t = (1-\rho^t)^2/(1-\rho)^2$. Using the law of iterated
expectations gives
\begin{equation*}
    \EE \exp\left(\sum_{t=0}^{n-1} Z_t \right)
    = \exp\left(
      (n+1)\mu + \frac{(1-\rho^{n+1})^2
        \sigma_\epsilon^2}{2(1-\rho)^2(1-\rho^2)} +
      \frac{\sigma_\epsilon^2}{2}\sum_{t=1}^n m_t  
    \right).
\end{equation*}
Since $m_t \to 1/(1-\rho)^2$, $\sum m_t/n \to 1/(1-\rho)^2$. Therefore,
\begin{equation}\label{eq:geomean2}
    \lim_{n\to\infty} \left( \EE \prod_{t=0}^{n-1}\beta_t \right)^{1/n}
    = \exp \left(
      \alpha \mu + \frac{\alpha^2\sigma_\epsilon^2}{2(1-\rho)^2}
    \right).
\end{equation}
Setting $\alpha=1$ gives \eqref{eq:geomean}. Setting $\mu = \log(b)$ and
$\alpha = 1/(1-1/\psi)$ gives \eqref{eq:EZ-r}.

\subsection{Necessity}
\label{ss:nec}

In many settings, the eventual discounting condition cannot be weakened
without violating finite lifetime values.  Here we briefly illustrate this point, 
using the connection to spectral radii provided in Proposition~\ref{p:0}.

Consider a standard dynamic program with lifetime rewards 
$\EE \sum_{t \geq 0} \beta^t \pi_t$ given
constant $\beta$ and reward flow $\{\pi_t\}$.
  In this setting, $\beta < 1$
cannot be relaxed without imposing specific conditions on rewards.  For
example, if there are constants $0 < a \leq b$ such that the process
$\{\pi_t\}$ satisfies  $a \leq \pi_t \leq b$
for all $t$, then we clearly have\footnote{The equivalence in \eqref{eq:es} is easy to
    see because, by the Monotone Convergence Theorem, we have $\EE \sum_{t
    \geq 0} \beta^t \pi_t = \sum_{t \geq 0} \beta^t \EE \pi_t$ and, moreover, $0 <
a \leq \EE \pi_t \leq b$.}
\begin{equation}
    \label{eq:es}
    \EE \sum_{t \geq 0} \beta^t \pi_t < \infty \text{ if and only if } \beta < 1.
\end{equation}
Eventual discounting has the same distinction once we replace the
constant $\beta$ with a process $\{ \beta_t
\}$ under standard regularity conditions.  For example, if $\ZZ$ is compact
and $\beta_t = \beta(Z_t)$ for some $\beta \in bc\ZZ$ and $Q$-Markov process $\{Z_t\}$,
then
\begin{equation}
    \label{eq:est}
    \EE_z \sum_{t \geq 0} \prod_{i=0}^{t-1} \beta_i  \, \pi_t < \infty
    \text{ if and only if } 
    r(L_\beta) < 1.
\end{equation}
To see this, suppose first that $r(L_\beta) < 1$.  Since $\pi_t \leq b$, we have
\begin{equation*}
    \EE_z \sum_{t \geq 0} \prod_{i=0}^{t-1} \beta_i  \, \pi_t
    \leq b \sum_{t \geq 0} \EE_z \prod_{i=0}^{t-1} \beta_i 
    \leq b \sum_{t \geq 0}  \sup_z \EE_z \prod_{i=0}^{t-1} \beta_i
    = b \sum_{t\geq 0} r_t^\beta.
\end{equation*}
By Cauchy's root convergence criterion, the sum $\sum_{t \geq 0} r^\beta_t$ will be finite
whenever $\limsup_{t \to \infty} (r^\beta_t)^{1/t} < 1$.  This holds
when $r(L_\beta) < 1$ by Proposition~\ref{p:0}.

Now suppose instead that $r(L_\beta) \geq 1$.  By compactness of
$L_\beta$, positivity of the function $\beta$ from Assumption~\ref{a:con}
and the Krein--Rutman Theorem (see, e.g., Theorem 1.2 in \cite{du2006order}), there exists a
positive function $e \in bc\ZZ$ such that $L_\beta e = r(L_\beta) e$.
Choosing $\gamma > 0$ such that $\gamma e \leq 1$, we have
\begin{equation*}
    \EE_z \sum_{t \geq 0} \prod_{i=0}^{t-1} \beta_i  \, \pi_t
    \geq a \gamma \sum_{t \geq 0} L_\beta^t e(z)
    = a \gamma \sum_{t \geq 0} r(L_\beta)^t e(z)
\end{equation*}
when $Z_0 = z$.
Since $e > 0$ and $r(L_\beta) \geq 1$, the sum diverges
to infinity.

\bibliographystyle{ecta}

\bibliography{sdd}

\end{document}